\documentclass[a4paper]{article}

\usepackage[utf8]{inputenc}
\usepackage[top=10mm]{geometry}
\usepackage{booktabs}
\usepackage[flushleft]{threeparttable}
\usepackage{authblk}
\usepackage{bm}
\usepackage{amsmath}
\usepackage{amssymb}
\usepackage{graphicx}
\usepackage[square,sort,numbers]{natbib}
\title{Meson-baryon scattering to one-loop order in heavy baryon chiral perturbation theory}

\author[1,2]{Bo-Lin Huang \thanks{bolin.huang@foxmail.com}}
\author[1]{Jin-Sheng Zhang\thanks{zjs19920501@163.com}}
\author[1]{Yun-De Li\thanks{deceased}}
\author[2]{Norbert Kaiser\thanks{nkaiser@ph.tum.de}}

\affil[1]{\normalsize Department of Physics, Yunnan University, Kunming 650091, China}
\affil[2]{\normalsize Physik Department, Technische Universit\"{a}t M\"{u}nchen, D-85747 Garching, Germany}

\date{\today}

\begin{document}
\maketitle

\begin{abstract}
We calculate the T-matrices of pseudoscalar meson octet-baryon scattering to one-loop order in SU(3) heavy baryon chiral perturbation theory. The pertinent combinations of low-energy constants are determined by fitting to phase shifts of $\pi N$ and $K N$ scattering and the corresponding data. By using these low-energy constants, we obtain the strong phase shift difference of $\pi \Lambda$ scattering at the $\Xi$ mass, $\delta_P-\delta_S=(8.8 \pm 0.2)$ degrees, in agreement with experimental results. We find that the phase shifts in the $S_{01}$($\pi \Sigma$), $S_{01}$($K \Xi$), $S_{01}$($\overline{K} N$), and $S_{11}$($\overline{K} \Sigma$) waves are so strong that resonances may be generated dynamically in all these channels. We also predict the scattering lengths and make comparisons with the results obtained from the threshold T-matrices in heavy baryon chiral perturbation theory and the method of covariant infrared regularization. The issue of convergence is also discussed in detail.
\begin{description}
\item[PACS numbers:]
13.75.Jz,12.39.Fe,12.38.Bx
\item[Keywords:]
Chiral perturbation theory, meson-baryon scattering
\end{description}
\end{abstract}

\section{Introduction}
As it is well known, quantum chromodynamics (QCD) becomes nonperturbative at low energies, and thus it is very difficult to use perturbative methods to calculate low-energy hadronic processes. For treating this problem, effective field theory (EFT) is introduced as a substitute for low-energy QCD. The EFT is formulated in terms of the most general Lagrangian consistent with the chiral symmetry of QCD, as well as the other continuous and discrete symmetries. In the EFT, the degrees of freedom are no longer quarks and gluons, but rather hadrons, i.e. pions, kaons, eta-mesons, and baryons. The corresponding field theoretical formalism is called chiral perturbation theory (ChPT) \cite{wein1979,sche2012}. The heavy baryon chiral perturbation theory (HB$\chi$PT) has been proposed and developed to solve the power-counting problem which arises from the nonvanishing baryon mass in the chiral limit \cite{jenk1991,bern1992}. Although relativistic approaches (such as infrared regularization \cite{bech1999} and the extended on-mass-shell scheme \cite{gege1999,fuch2003}) have made substantial progress in many aspects \cite{schi2007,geng2008,mart2010,chen2012,ren2012,alar2013}, HB$\chi$PT is still a reasonable and powerful tool for the study of meson-baryon scattering processes.  Dark matter interactions with mesons or nucleons \cite{fan2010,bish2017} and the breaking of Lorentz and CPT symmetry \cite{noor2016} can also be investigated in HB$\chi$PT.

Over the years, SU(2) HB$\chi$PT has been widely used to investigate the low-energy processes of pions and nucleons and has achieved many successes \cite{fett1998,fett2000,kreb2012,ente2015}. For processes involving kaons or hyperons, one has to use three-flavor chiral dynamics.  In a previous paper \cite{huan2015} we have investigated the $KN$ and $\overline{K}N$ scattering to one-loop order in SU(3) HB$\chi$PT by fitting to partial-wave phase shifts of $KN$ scattering and obtained reasonable results. In this paper, we will extend this approach to predictions for pseudoscalar meson octet-baryon scattering in all channels by fitting to partial-wave phase shifts of elastic $\pi N$ and $KN$ scattering simultaneously. Note that the predictions for the meson-baryon scattering lengths in SU(3) HB$\chi$PT have been given in detail  and reasonable results have been obtained in Refs.~\cite{kais2001,liu20071,liu20072,liu2011,liu2012}. Furthermore, meson-baryon scattering lengths have also been calculated in covariant baryon chiral perturbation theory using the method of infrared regularization in Ref.~\cite{mai2009}. However, our study is not only concerned with scattering lengths but also with partial-wave phase shifts. One can obtain much more information from partial-wave phase shifts than scattering lengths in meson-baryon scattering because the complete information about a scattering process in the physical region is contained in partial-wave phase shifts. Therefore, our extension of the calculations in HB$\chi$PT are interesting.

In Sec.~\ref{lagrangian}, the Lagrangians involved in the calculations up to one-loop order are presented in detail. In Sec.~\ref{tmatrix}, we present the T-matrices for the meson-baryon scattering processes order by order and channel by channel. In Sec.~\ref{phase}, we outline how to calculate phase shifts and scattering lengths. Section \ref{results} is devoted to the presentation and discussion of our results and it includes also a brief summary.
\section{Chiral Lagrangian}
\label{lagrangian}
Our calculation of meson-baryon scattering is based on the SU(3) effective chiral Lagrangian in the heavy baryon formulation,
\begin{align}
\label{eq1}
\mathcal{L}=\mathcal{L}_{\phi\phi}+\mathcal{L}_{\phi B}.
\end{align}
The traceless Hermitian $3\times 3$ matrices $\phi$ and $B$ include the pseudoscalar Goldstone boson fields ($\pi,K,\overline{K},\eta$) and the octet baryon fields ($N, \Lambda, \Sigma, \Xi$), respectively. The lowest-order  SU(3) chiral Lagrangians for meson-meson and meson-baryon interactions take the form \cite{bora1997}
\begin{align}
\label{eq2}
\mathcal{L}_{\phi\phi}^{(2)}=\frac{f^{2}}{4}\text{tr}(u_{\mu}u^{\mu}+\chi_{+}),
\end{align}
\begin{align}
\label{eq3}
 \mathcal{L}_{\phi B}^{(1)}=\text{tr}(i\overline{B}[v\cdot D,B])+D\,\text{tr}(\overline{B}S_{\mu}\{u^{\mu},B\})+F\,\text{tr}(\overline{B}S_{\mu}[u^{\mu},B]),
\end{align}
where $D_{\mu}$ denotes the chiral covariant derivative
\begin{align}
\label{eq4}
[D_{\mu},B]=\partial_{\mu}B+[\Gamma_{\mu},B],
\end{align}
and $S_{\mu}$ is the covariant spin operator. In practice one works with the Pauli spin matrices $\vec{\sigma}$, such that:
\begin{align}
\label{eq5}
S^{\mu}=\Big(0,\frac{\vec{\sigma}}{2}\Big).
\end{align}
The chiral connection $\Gamma^{\mu}=[\xi^{\dag},\partial^{\mu}\xi]/2$ and the axial vector quantity $u^{\mu}=i\{\xi^{\dag},\partial^{\mu}\xi\}$ contain an even and odd number of meson fields, respectively. The SU(3) matrix $U=\xi^{2}=\text{exp}(i\phi/f)$ collects the pseudoscalar Goldstone boson fields. The parameter $f$ is the pseudoscalar decay constant in the chiral limit. The axial vector coupling constants $D$ and $F$ can be determined in fits to semileptonic hyperon decays \cite{bora1999}. The quantity $\chi_{+}=\xi^{\dag}\chi\xi^{\dag}+\xi\chi\xi$ with $\chi=\text{diag}(m_{\pi}^{2},m_{\pi}^{2},2m_{K}^{2}-m_{\pi}^{2})$ introduces explicit chiral symmetry breaking terms. The complete heavy-baryon Lagrangian  at next-to-leading order splits up into two parts
\begin{align}
\label{eq6}
\mathcal{L}_{\phi B}^{(2)}=\mathcal{L}_{\phi B}^{(2,1/M_{0})}+\mathcal{L}_{\phi B}^{(2,\text{ct})},
\end{align}
where $\mathcal{L}_{\phi B}^{(2,1/M_{0})}$ denotes $1/M_{0}$ corrections of dimension two with fixed coefficients and it stems from the $1/M_{0}$ expansion of the original relativistic leading-order Lagrangian $\mathcal{L}_{\phi B}^{(1)}$ \cite{bora1997}. The pertinent terms read
\begin{align}
\label{eq7}
\mathcal{L}_{\phi B}^{(2,1/M_{0})}=\,\,&\frac{D^{2}-3F^{2}}{24M_{0}}\text{tr}(\overline{B}[v\cdot u,[v\cdot u,B]])
-\frac{D^{2}}{12M_{0}}\text{tr}(\overline{B}B)\text{tr}(v\cdot u\,\, v\cdot u)\nonumber\\
&-\frac{DF}{4M_{0}}\text{tr}(\overline{B}[v\cdot u,\{v\cdot u,B\}])
-\frac{1}{2M_{0}}\text{tr}(\overline{B}[D_{\mu},[D^{\mu},B]])\nonumber\\
&+\frac{1}{2M_{0}}\text{tr}(\overline{B}[v\cdot D,[v\cdot D,B]])
-\frac{iD}{2M_{0}}\text{tr}(\overline{B}S_{\mu}[D^{\mu},\{v\cdot u,B\}])\nonumber\\
&-\frac{iF}{2M_{0}}\text{tr}(\overline{B}S_{\mu}[D^{\mu},[v\cdot u,B]])
-\frac{iF}{2M_{0}}\text{tr}(\overline{B}S_{\mu}[v\cdot u,[D^{\mu},B]])\nonumber\\
&-\frac{iD}{2M_{0}}\text{tr}(\overline{B}S_{\mu}\{v\cdot u,[D^{\mu},B]\}),
\end{align}
where $M_{0}$ denotes the baryon mass in the chiral limit. The remaining meson-baryon Lagrangian $\mathcal{L}_{\phi B}^{(2,\text{ct})}$ involves new low-energy constants (LECs) and it can be obtained from the relativistic Lagrangian in Ref.~\cite{olle2006}. The pertinent interaction terms read
\begin{align}
\label{eq8}
\mathcal{L}_{\phi B}^{(2,\text{ct})}=&\,\,b_{D}\,\text{tr}(\overline{B}\{\chi_{+},B\})+b_{F}\,\text{tr}(\overline{B}[\chi_{+},B])
+b_{0}\,\text{tr}(\overline{B}B)\text{tr}(\chi_{+})+b_{1}\,\text{tr}(\overline{B}\{u^{\mu}u_{\mu},B\})\nonumber\\
&+b_{2}\,\text{tr}(\overline{B}[u^{\mu}u_{\mu},B])+b_{3}\,\text{tr}(\overline{B}B)\text{tr}(u^{\mu}u_{\mu})
+b_{4}\,\text{tr}(\overline{B}u^{\mu})\text{tr}(u_{\mu}B)+b_{5}\,\text{tr}(\overline{B}\{v\cdot u\,\,v\cdot u,B\})\nonumber\\
&+b_{6}\,\text{tr}(\overline{B}[v\cdot u\,\,v\cdot u,B])+b_{7}\,\text{tr}(\overline{B}B)\text{tr}(v\cdot u\,\,v\cdot u)
+b_{8}\,\text{tr}(\overline{B}v\cdot u )\text{tr}( v\cdot u B)\nonumber\\
&+b_{9}\,\text{tr}(\overline{B}\{[u^{\mu},u^{\nu}],[S_{\mu},S_{\nu}]B\})
+b_{10}\,\text{tr}(\overline{B}[[u^{\mu},u^{\nu}],[S_{\mu},S_{\nu}]B])\nonumber\\
&+b_{11}\,\text{tr}(\overline{B}u^{\mu})\text{tr}(u^{\nu}[S_{\mu},S_{\nu}]B).
\end{align}
The first three terms proportional to the LECs $b_{D,F,0}$ give rise to explicit chiral symmetry breaking. Note that all LECs $b_{i}$ have dimension $\text{mass}^{-1}$.

\section{T-matrices for meson-baryon scattering}
\label{tmatrix}
We are considering in this work only elastic meson-baryon scattering processes $M(\bm{q})+B(\bm{-q}) \rightarrow M(\bm{q}')+B(\bm{-q}')$ in the center-of-mass system with $|\bm{q}|=|\bm{q}'|=q$. The corresponding T-matrix takes the following form:
\begin{align}
\label{eq9}
 T_{MB}^{(I)}=&V_{MB}^{(I)}(q)+i\bm{\sigma}\cdot(\bm{q}'\times\bm{q})W_{MB}^{(I)}(q),
\end{align}
where $I$ denotes the total isospin of the meson-baryon system. Furthermore, $V_{MB}^{(I)}(q)$ refers to the non-spin-flip meson-baryon amplitude and $W_{MB}^{(I)}(q)$ refers to the spin-flip meson-baryon amplitude. In the following subsections we calculate the T-matrices order by order for every channel and specify them by giving the pair $T_{MB}^{(I)}=\{V_{MB}^{(I)},\,W_{MB}^{(I)}\}$. The velocity four-vector is chosen as $v^{\mu}=(1,0,0,0)$ throughout this paper.

\begin{figure}[t]
\centering
\includegraphics[height=5cm,width=8cm]{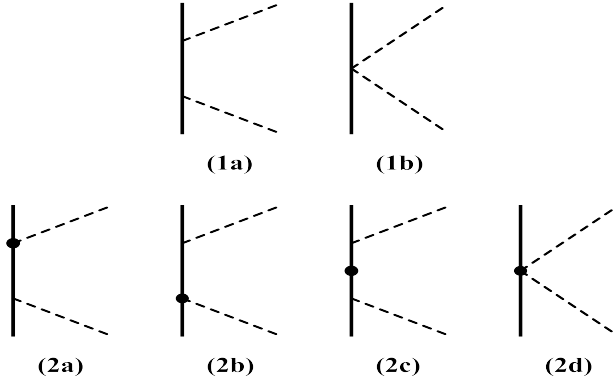}
\caption{\label{fig:treefeynman}Tree diagrams contributing at first and second chiral order. Dashed lines represent Goldstone bosons and solid lines represent octet baryons. The heavy dots refer to vertices from $\mathcal{L}_{\phi B}^{(2)}$. Diagrams with crossed meson lines are not shown. } 
\end{figure}

\subsection{Leading order amplitudes}
\label{LO}
For elastic meson-baryon scattering, the leading order $\mathcal{O}(q)$ amplitudes resulting from diagrams (1a) and (1b) in Fig.~\ref{fig:treefeynman} (and their crossed partners) read
\begin{align}
\label{eq10}
T_{\pi N}^{(3/2,\text{LO})}=\Bigg\{\frac{-w_{\pi}^2+q^2 z (D+F)^2}{2 f_{\pi}^2 w_{\pi}},\,-\frac{(D+F)^2}{2 f_{\pi}^2 w_{\pi}}\Bigg\},
\end{align}
\begin{align}
\label{eq11}
T_{\pi N}^{(1/2,\text{LO})}=\Bigg\{\frac{w_{\pi}^2-q^2 z (D+F)^2}{f_{\pi}^2 w_{\pi}},\,-\frac{(D+F)^2}{2f_{\pi}^2 w_{\pi}}\Bigg\},
\end{align}
\begin{align}
\label{eq12}
T_{\pi \Sigma}^{(2,\text{LO})}=\Bigg\{\frac{-3 w_{\pi}^2+q^2 z(D^2+3F^2)}{3 f_{\pi}^2 w_{\pi}},\,-\frac{D^2+3F^2}{3 f_{\pi}^2 w_{\pi}}\Bigg\},
\end{align}
\begin{align}
\label{eq13}
T_{\pi \Sigma}^{(1,\text{LO})}=\Bigg\{\frac{3 w_{\pi}^2-q^2 z (D^2+3 F^2)}{3 f_{\pi}^2 w_{\pi}},\,\frac{D^2-9F^2}{3 f_{\pi}^2 w_{\pi}}\Bigg\},
\end{align}
\begin{align}
\label{eq14}
T_{\pi \Sigma}^{(0,\text{LO})}=\Bigg\{\frac{6w_{\pi}^2-2 q^2 z (D^2+3 F^2)}{3 f_{\pi}^2 w_{\pi}},\frac{-4D^2+6F^2}{3 f_{\pi}^2 w_{\pi}}\Bigg\},
\end{align}
\begin{align}
\label{eq15}
T_{\pi \Xi}^{(3/2,\text{LO})}=\Bigg\{\frac{-w_{\pi}^2+ q^2 z (D-F)^2}{2 f_{\pi}^2 w_{\pi}},\,-\frac{(D-F)^2}{2 f_{\pi}^2 w_{\pi}}\Bigg\},
\end{align}
\begin{align}
\label{eq16}
T_{\pi \Xi}^{(1/2,\text{LO})}=\Bigg\{\frac{w_{\pi}^2- q^2 z (D-F)^2}{f_{\pi}^2 w_{\pi}},\,-\frac{(D-F)^2}{2f_{\pi}^2 w_{\pi}}\Bigg\},
\end{align}
\begin{align}
\label{eq17}
T_{\pi \Lambda}^{(\text{LO})}=\Bigg\{0,-\frac{2D^2}{3f_{\pi}^2 w_{\pi}}\Bigg\},
\end{align}
\begin{align}
\label{eq18}
T_{KN}^{(1,\text{LO})}=\Bigg\{\frac{-3w_{K}^2+q^{2}z(D^{2}+3F^{2})}{3f_{K}^{2}w_{K}},\,-\frac{D^{2}+3F^{2}}{3f_{K}^{2}w_{K}}\Bigg\},
\end{align}
\begin{align}
\label{eq19}
T_{KN}^{(0,\text{LO})}=\Bigg\{\frac{2q^{2}zD(D-3F)}{3f_{K}^{2}w_{K}},\,-\frac{2D(D-3F)}{3f_{K}^{2}w_{K}}\Bigg\},
\end{align}
\begin{align}
\label{eq20}
T_{\overline{K}N}^{(1,\text{LO})}=\Bigg\{\frac{w_{K}^2-q^{2}z(D-F)^2}{2f_{K}^{2}w_{K}},-\frac{(D-F)^{2}}{2f_{K}^{2}w_{K}}\Bigg\},
\end{align}
\begin{align}
\label{eq21}
T_{\overline{K}N}^{(0,\text{LO})}=\Bigg\{\frac{9w_{K}^2-q^{2}z(D+3F)^2}{6f_{K}^{2}w_{K}},\,-\frac{(D+3F)^{2}}{6f_{K}^{2}w_{K}}\Bigg\},
\end{align}
\begin{align}
\label{eq22}
T_{K \Sigma}^{(3/2,\text{LO})}=\Bigg\{\frac{-w_{K}^2+q^2 z (D+F)^2}{2f_{K}^2 w_{K}},\,-\frac{ (D+F)^2}{2f_{K}^2 w_{K}}\Bigg\},
\end{align}
\begin{align}
\label{eq23}
T_{K \Sigma}^{(1/2,\text{LO})}=\Bigg\{\frac{w_{K}^2-q^2 z (D^2-DF+F^2)}{f_{K}^2 w_{K}},
-\frac{D^2-4DF+F^2}{2f_{K}^2 w_{K}}\Bigg\},
\end{align}
\begin{align}
\label{eq24}
T_{\overline{K} \Sigma}^{(3/2,\text{LO})}=\Bigg\{\frac{-w_{K}^2+q^2 z (D-F)^2}{2 f_{K}^2 w_{K}},\,-\frac{(D-F)^2}{2f_{K}^2 w_{K}}\Bigg\},
\end{align}
\begin{align}
\label{eq25}
T_{\overline{K} \Sigma}^{(1/2,\text{LO})}=\Bigg\{\frac{w_{K}^2- q^2 z (D^2+DF+F^2)}{f_{K}^2 w_{K}},-\frac{D^2+4DF+F^2}{2f_{K}^2 w_{K}}\Bigg\},
\end{align}
\begin{align}
\label{eq26}
T_{K \Xi}^{(1,\text{LO})}=\Bigg\{\frac{w_{K}^2- q^2 z (D+F)^2}{2f_{K}^2 w_{K}},\,-\frac{(D+F)^2}{2f_{K}^2 w_{K}}\Bigg\},
\end{align}
\begin{align}
\label{eq27}
T_{K \Xi}^{(0,\text{LO})}=\Bigg\{\frac{9w_{K}^2-q^2 z (D-3 F)^2}{6 f_{K}^2 w_{K}},\,-\frac{(D-3F)^2}{6 f_{K}^2 w_{K}}\Bigg\},
\end{align}
\begin{align}
\label{eq28}
T_{\overline{K} \Xi}^{(1,\text{LO})}=\Bigg\{\frac{-3w_{K}^2+ q^2 z (D^2+3 F^2)}{3 f_{K}^2 w_{K}},\,-\frac{D^2+3F^2}{3 f_{K}^2 w_{K}}\Bigg\},
\end{align}
\begin{align}
\label{eq29}
T_{\overline{K} \Xi}^{(0,\text{LO})}=\Bigg\{\frac{2 q^2 z D(D+3 F)}{3 f_{K}^2 w_{K}},\,-\frac{2D(D+3F)}{3 f_{K}^2 w_{K}}\Bigg\},
\end{align}
\begin{align}
\label{eq30}
T_{K\Lambda}^{(\text{LO})}=\Bigg\{-\frac{ q^2 z D F}{f_{K}^2 w_{K}},\,-\frac{D^2+9F^2}{6f_{K}^2 w_{K}}\Bigg\},
\end{align}
\begin{align}
\label{eq31}
T_{\overline{K} \Lambda}^{(\text{LO})}=\Bigg\{\frac{ q^2 z D F}{f_{K}^2 w_{K}},\,-\frac{D^2+9F^2}{6f_{K}^2 w_{K}}\Bigg\},
\end{align}
\begin{align}
\label{eq32}
T_{\eta N}^{(\text{LO})}=\Bigg\{0,\,-\frac{(D-3F)^2}{6f_{\eta}^2 w_{\eta}}\Bigg\},
\end{align}
\begin{align}
\label{eq33}
T_{\eta \Sigma}^{(\text{LO})}=\Bigg\{0,\,-\frac{2D^2}{3f_{\eta}^2 w_{\eta}}\Bigg\},
\end{align}
\begin{align}
\label{eq34}
T_{\eta \Xi}^{(\text{LO})}=\Bigg\{0,\,-\frac{(D+3F)^2}{6f_{\eta}^2 w_{\eta}}\Bigg\},
\end{align}
\begin{align}
\label{eq35}
T_{\eta \Lambda}^{(\text{LO})}=\Bigg\{0,\,-\frac{2D^2}{3f_{\eta}^2 w_{\eta}}\Bigg\},
\end{align}
where $w_{\pi,K,\eta}=(m_{\pi,K,\eta}^{2}+q^{2})^{1/2}$ denotes the center-of-mass energy of the pion, kaon and eta, respectively, and $z=\text{cos}(\theta)$ is the cosine of the angle $\theta$ between $\bm{q}$ and $\bm{q}'$. We take the renormalized (physical) decay constants $f_{\pi,K,\eta}$ instead of $f$ (the chiral limit value). Note that, in the channels with an isoscalar $\eta$-meson or $\Lambda$-hyperon, the total isospin $I$ is unique and does not need to be specified.  

\subsection{Next-to-leading order amplitudes}
At next-to-leading order $\mathcal{O}(q^{2})$ one has the contributions from the diagrams in the second row of Fig.~\ref{fig:treefeynman} (including crossed diagrams), which involve vertices from the Lagrangians $\mathcal{L}_{\phi B}^{(2,1/M_{0})}$ and $\mathcal{L}_{\phi B}^{(2,\text{ct})}$. First, for the vertices from $\mathcal{L}_{\phi B}^{(2,1/M_{0})}$, the amplitudes have the generic polynomial form,
\begin{align}
\label{eq36}
V^{(1/M_0)}_{\pi,K,\eta}[a,b,c,d]=a\frac{1}{M_0 w_{\pi,K,\eta}^2 f_{\pi,K,\eta}^2}(bq^4+cq^2w_{\pi,K,\eta}^2+dw_{\pi,K,\eta}^4),
\end{align}
\begin{align}
\label{eq37}
W^{(1/M_0)}_{\pi,K,\eta}[e,f,g]=e\frac{1}{M_0 w_{\pi,K,\eta}^2 f_{\pi,K,\eta}^2}(fq^2+gw_{\pi,K,\eta}^2).
\end{align}
We specify only the coefficients $a, b, c, d$ for $V$ and $e, f, g$ for $W$ by listing them in square brackets:
\begin{align}
\label{eq38}
T_{\pi N}^{(3/2,1/M_0)}=&\Bigg\{V^{(1/M_0)}_{\pi}\Bigg[\frac{(D+F)^2}{4},-2z(1+z),\frac{-1-z}{(D+F)^2}+2(2+z),-1\Bigg],
\nonumber\\
&W^{(1/M_0)}_{\pi}\Bigg[\frac{(D+F)^2}{2},(1+z),-1\Bigg]\Bigg\},
\end{align}
\begin{align}
\label{eq39}
T_{\pi N}^{(1/2,1/M_0)}=&\Bigg\{V^{(1/M_0)}_{\pi}\Bigg[\frac{(D+F)^2}{4},z(1+z),\frac{2(1+z)}{(D+F)^2}-2(1+2z),-1\Bigg],
\nonumber\\
&W^{(1/M_0)}_{\pi}\Bigg[\frac{(D+F)^2}{4},-1-z,-2\Bigg]\Bigg\},
\end{align}
\begin{align}
\label{eq40}
T_{\pi \Sigma}^{(2,1/M_0)}=&\Bigg\{V^{(1/M_0)}_{\pi}\Bigg[\frac{D^2+3F^2}{6},-2z(1+z),\frac{-3(1+z)}{D^2+3F^2}+2(2+z),-1\Bigg],\nonumber\\
&W^{(1/M_0)}_{\pi}\Bigg[\frac{D^2+3F^2}{6},2(1+z),-2\Bigg]\Bigg\},
\end{align}
\begin{align}
\label{eq41}
T_{\pi \Sigma}^{(1,1/M_0)}=&\Bigg\{V^{(1/M_0)}_{\pi}\Bigg[\frac{D^2+3F^2}{6},2z(1+z),\frac{3(1+z)}{D^2+3F^2}-2(2+z),\frac{D^2-9F^2}{D^2+3F^2}\Bigg],\nonumber\\
&W^{(1/M_0)}_{\pi}\Bigg[\frac{D^2+3F^2}{3},-(1+z),1\Bigg]\Bigg\},
\end{align}
\begin{align}
\label{eq42}
T_{\pi \Sigma}^{(0,1/M_0)}=&\Bigg\{V^{(1/M_0)}_{\pi}\Bigg[\frac{D^2+3F^2}{3},-z(1+z),\frac{3(1+z)}{D^2+3F^2}-2(z-1),\frac{-2D^2+3F^2}{D^2+3F^2}\Bigg],\nonumber\\
&W^{(1/M_0)}_{\pi}\Bigg[\frac{D^2+3F^2}{3},1+z,-4)\Bigg\},
\end{align}
\begin{align}
\label{eq43}
T_{\pi \Xi}^{(3/2,1/M_0)}=&\Bigg\{V^{(1/M_0)}_{\pi}\Bigg[\frac{(D-F)^2}{4},-2z(1+z),\frac{-(1+z)}{(D-F)^2}+2(2+z),-1\Bigg],\nonumber\\
&W^{(1/M_0)}_{\pi}\Bigg[\frac{(D-F)^2}{4},2(1+z),-2\Bigg]\Bigg\},
\end{align}
\begin{align}
\label{eq44}
T_{\pi \Xi}^{(1/2,1/M_0)}=&\Bigg\{V^{(1/M_0)}_{\pi}\Bigg[\frac{(D-F)^2}{4},z(1+z),\frac{2(1+z)}{(D-F)^2}-2(1+2z),-1\Bigg],\nonumber\\
&W^{(1/M_0)}_{\pi}\Bigg[\frac{(D-F)^2}{4},-1-z,-2\Bigg]\Bigg\},
\end{align}
\begin{align}
\label{eq45}
T_{\pi\Lambda}^{(1/M_0)}=&\Bigg\{-\frac{D^2w_{\pi}^2}{3M_0f_{\pi}^2},\,0\Bigg\},
\end{align}
\begin{align}
\label{eq46}
T_{K N}^{(1,1/M_0)}=&\Bigg\{V^{(1/M_0)}_{K}\Bigg[\frac{D^2+3F^2}{6},-2z(1+z),\frac{-3(1+z)}{D^2+3F^2}+2(2+z),-1\Bigg],\nonumber\\
&W^{(1/M_0)}_{K}\Bigg[\frac{D^2+3F^2}{6},2(1+z),-2\Bigg]\Bigg\},
\end{align}
\begin{align}
\label{eq47}
T_{K N}^{(0,1/M_0)}=&\Bigg\{V^{(1/M_0)}_{K}\Bigg[\frac{D(D-3F)}{3},-2z(1+z),2(2+z),-1\Bigg],\nonumber\\
&W^{(1/M_0)}_{K}\Bigg[\frac{D(D-3F)}{3},2(1+z),-2\Bigg]\Bigg\},
\end{align}
\begin{align}
\label{eq48}
T_{\overline{K} N}^{(1,1/M_0)}=&\Bigg\{V^{(1/M_0)}_{K}\Bigg[\frac{(D-F)^2}{4},0,\frac{1+z}{(D-F)^2}-2z,-1\Bigg],\,W^{(1/M_0)}_{K}\Bigg[\frac{(D-F)^2}{2},0,-1\Bigg]\Bigg\},
\end{align}
\begin{align}
\label{eq49}
T_{\overline{K} N}^{(0,1/M_0)}=&\Bigg\{V^{(1/M_0)}_{K}\Bigg[\frac{(D+3F)^2}{12},0,\frac{9(1+z)}{(D+3F)^2}-2z,-1\Bigg],\,W^{(1/M_0)}_{K}\Bigg[\frac{(D+3F)^2}{6},0,-1\Bigg]\Bigg\},
\end{align}
\begin{align}
\label{eq50}
T_{K\Sigma}^{(3/2,1/M_0)}=&\Bigg\{V^{(1/M_0)}_{K}\Bigg[\frac{(D+F)^2}{4},-2z(1+z),\frac{-1-z}{(D+F)^2}+2(z+2),-1\Bigg],\nonumber\\
&W^{(1/M_0)}_{K}\Bigg[\frac{(D+F)^2}{4},2(1+z),-2)\Bigg]\Bigg\},
\end{align}
\begin{align}
\label{eq51}
T_{K\Sigma}^{(1/2,1/M_0)}=&\Bigg\{V^{(1/M_0)}_{K}\Bigg[\frac{(D+F)^2}{4},z(1+z),\frac{2-2z+4z(D^2-DF+F^2)}{(D+F)^2}-2,-\frac{D^2-4DF+F^2}{(D+F)^2}\Bigg],\nonumber\\
&W^{(1/M_0)}_{K}\Bigg[\frac{(D+F)^2}{4},-1-z,-\frac{2(D^2-4DF+F^2)}{(D+F)^2}\Bigg]\Bigg\},
\end{align}
\begin{align}
\label{eq52}
T_{\overline{K}\Sigma}^{(3/2,1/M_0)}=&\Bigg\{V^{(1/M_0)}_{K}\Bigg[\frac{(D-F)^2}{4},-2z(1+z),\frac{-1-z}{(D-F)^2}+2(z+2),-1\Bigg],\nonumber\\
&W^{(1/M_0)}_{K}\Bigg[\frac{(D-F)^2}{4},2(1+z),-2\Bigg]\Bigg\},
\end{align}
\begin{align}
\label{eq53}
T_{\overline{K}\Sigma}^{(1/2,1/M_0)}=&\Bigg\{V^{(1/M_0)}_{K}\Bigg[\frac{(D-F)^2}{4},z(1+z),\frac{2+2z-4z(D^2+DF+F^2)}{(D-F)^2}-2,-\frac{D^2+4DF+F^2}{(D-F)^2}\Bigg],\nonumber \\
&W^{(1/M_0)}_{K}\Bigg[\frac{(D-F)^2}{4},-1-z,-\frac{2(D^2+DF+F^2)}{(D-F)^2}\Bigg]\Bigg\},
\end{align}
\begin{align}
\label{eq54}
T_{K\Xi}^{(1,1/M_0)}=\Bigg\{V^{(1/M_0)}_{K}\Bigg[\frac{(D+F)^2}{4},0,\frac{1+z}{(D+F)^2}-2z,-1\Bigg],\,W^{(1/M_0)}_{K}\Bigg[\frac{(D+F)^2}{2},0,-1\Bigg]\Bigg\},
\end{align}
\begin{align}
\label{eq55}
T_{K\Xi}^{(0,1/M_0)}=\Bigg\{V^{(1/M_0)}_{K}\Bigg[\frac{(D-3F)^2}{12},0,\frac{9(1+z)}{(D-3F)^2}-2z,-1\Bigg],\,W^{(1/M_0)}_{K}\Bigg[\frac{(D-3F)^2}{6},0,-1\Bigg]\Bigg\},
\end{align}
\begin{align}
\label{eq56}
T_{\overline{K}\Xi}^{(1,1/M_0)}=&\Bigg\{V^{(1/M_0)}_{K}\Bigg[\frac{D^2+3F^2}{6},-2z(1+z),-\frac{3(1+z)}{D^2+3F^2}+4+2z,-1\Bigg],\nonumber\\
&W^{(1/M_0)}_{K}\Bigg[\frac{D^2+3F^2}{6},2(1+z),-2\Bigg]\Bigg\},
\end{align}
\begin{align}
\label{eq57}
T_{\overline{K}\Xi}^{(0,1/M_0)}=&\Bigg\{V^{(1/M_0)}_{K}\Bigg[\frac{D(D+3F)}{3},-2z(1+z),2(2+z),-1\Bigg],\nonumber\\
&W^{(1/M_0)}_{K}\Bigg[\frac{D(D+3F)}{3},2(1+z),-2\Bigg]\Bigg\},
\end{align}
\begin{align}
\label{eq58}
T_{K\Lambda}^{(1/M_0)}=&\Bigg\{V^{(1/M_0)}_{K}\Bigg[-\frac{1}{12},z(1+z)(D-3F)^2,-2D^2-18F^2+12(1+z)DF,D^2+9F^2\Bigg],\nonumber\\
&W^{(1/M_0)}_{K}\Bigg[\frac{1}{12},(1+z)(D-3F)^2,-2(D^2+9F^2)\Bigg]\Bigg\},
\end{align}
\begin{align}
\label{eq59}
T_{\overline{K}\Lambda}^{(1/M_0)}=&\Bigg\{V^{(1/M_0)}_{K}\Bigg[-\frac{1}{12},z(1+z)(D+3F)^2,-2D^2-18F^2-12(1+z)DF,D^2+9F^2\Bigg],\nonumber\\
&W^{(1/M_0)}_{K}\Bigg[\frac{1}{12},(1+z)(D+3F)^2,-2(D^2+9F^2)\Bigg]\Bigg\},
\end{align}
\begin{align}
\label{eq60}
T_{\eta N}^{(1/M_0)}=\Bigg\{V^{(1/M_0)}_{\eta}\Bigg[\frac{(D-3F)^2}{12},(-1-z)z,2+z,-1\Bigg],\,W^{(1/M_0)}_{\eta}\Bigg[\frac{(D-3F)^2}{12},1+z,-1\Bigg]\Bigg\},
\end{align}
\begin{align}
\label{eq61}
T_{\eta \Sigma}^{(1/M_0)}=\Bigg\{-\frac{D^2w_{\eta}^2}{3f_{\eta}^2M_0},\,0\Bigg\},
\end{align}
\begin{align}
\label{eq62}
T_{\eta\Xi}^{(1/M_0)}=\Bigg\{V^{(1/M_0)}_{\eta}\Bigg[\frac{(D+3F)^2}{12},(-1-z)z,2+z,-1\Bigg],\,W^{(1/M_0)}_{\eta}\Bigg[\frac{(D+3F)^2}{12},1+z,-1\Bigg]\Bigg\},
\end{align}
\begin{align}
\label{eq63}
T_{\eta \Lambda}^{(1/M_0)}=\Bigg\{-\frac{D^2w_{\eta}^2}{3f_{\eta}^2M_0},\,0\Bigg\}.
\end{align}

For the explicit chiral symmetry breaking part of $\mathcal{L}_{\phi B}^{(2,\text{ct})}$, the amplitudes involving LECs read

\begin{align}
\label{eq64}
T_{\pi N}^{(3/2,\text{NLO})}=&\Bigg\{\frac{2}{f_\pi^2}[-(2b_0+b_D+b_F)m_\pi^2-C_1 zq^2+(C_1+C_2)w_\pi^2],\,-\frac{2}{f_\pi^2}C_3 \Bigg\},
\end{align}
\begin{align}
\label{eq65}
T_{\pi N}^{(1/2,\text{NLO})}=&\Bigg\{\frac{2}{f_\pi^2}[-(2b_0+b_D+b_F)m_\pi^2-C_1 zq^2+(C_1+C_2)w_\pi^2],\,\frac{4}{f_\pi^2}C_3\Bigg\},
\end{align}
\begin{align}
\label{eq66}
T_{K N}^{(1,\text{NLO})}=&\Bigg\{\frac{1}{f_K^2}[-4(b_0+b_D)m_K^2-2C_4zq^2+2(C_4+C_5)w_K^2],\,-\frac{1}{f_K^2}C_6\Bigg\},
\end{align}
\begin{align}
\label{eq67}
T_{K N}^{(0,\text{NLO})}=&\Bigg\{\frac{1}{f_K^2}[-4(b_0-b_F)m_K^2+2C_7zq^2-2(C_7+C_8)w_K^2],\,\frac{1}{f_K^2}C_9\Bigg\},
\end{align}
\begin{align}
\label{eq68}
T_{\overline{K} N}^{(1,\text{NLO})}=&\Bigg\{\frac{1}{f_K^2}[-2(2b_0+b_D-b_F)m_K^2-(C_4-C_7)zq^2+(C_4+C_5\nonumber\\
&-C_7-C_8)w_K^2],\,\frac{1}{2f_K^2}(C_6-C_9)\Bigg\},
\end{align}
\begin{align}
\label{eq69}
T_{\overline{K} N}^{(0,\text{NLO})}=&\Bigg\{\frac{1}{f_K^2}[-2(2b_0+3b_D+b_F)m_K^2-(3C_4+C_7)zq^2+(3C_4+C_7\nonumber\\
&+3C_5+C_8)w_K^2],\frac{1}{2f_K^2}(3C_6+C_9)\Bigg\},
\end{align}
\begin{align}
\label{eq70}
T_{\pi \Sigma}^{(2,\text{NLO})}=&\Bigg\{\frac{1}{f_\pi^2}[-4(b_0+b_D)m_\pi^2-2C_4 zq^2+2(C_4+C_5)w_\pi^2],\,-\frac{1}{f_\pi^2}C_6\Bigg\},
\end{align}
\begin{align}
\label{eq71}
T_{\pi \Sigma}^{(1,\text{NLO})}=&\Bigg\{-\frac{1}{f_\pi^2}[4(b_0+b_D)m_\pi^2+2(C_4-2b_4) zq^2-2(C_4-2b_4+C_5-2b_8)w_\pi^2],\nonumber\\
&\frac{1}{f_\pi^2}C_6\Bigg\},
\end{align}
\begin{align}
\label{eq72}
T_{\pi \Sigma}^{(0,\text{NLO})}=&\Bigg\{\frac{2}{f_\pi^2}[-2(b_0+b_D)m_\pi^2-(C_4+3b_4) zq^2+(C_4+3b_4+C_5+3b_8)w_\pi^2],\nonumber\\
&\frac{2}{f_\pi^2}C_6\Bigg\},
\end{align}
\begin{align}
\label{eq73}
T_{\pi \Xi}^{(3/2,\text{NLO})}=&\Bigg\{\frac{1}{f_\pi^2}[-2(2b_0+b_D-b_F)m_\pi^2-(C_4-C_7) zq^2+(C_4-C_7\nonumber\\
&+C_5-C_8)w_\pi^2],\,-\frac{1}{2f_\pi^2}(C_6-C_9)\Bigg\},
\end{align}
\begin{align}
\label{eq74}
T_{\pi \Xi}^{(1/2,\text{NLO})}=&\Bigg\{\frac{1}{f_\pi^2}[-2(2b_0+b_D-b_F)m_\pi^2-(C_4-C_7) zq^2+(C_4-C_7\nonumber\\
&+C_5-C_8)w_\pi^2],\,\frac{1}{f_\pi^2}(C_6-C_9)\Bigg\},
\end{align}
\begin{align}
\label{eq75}
T_{\pi \Lambda}^{(\text{NLO})}=\Bigg\{\frac{2}{3f_\pi^2}[-2(3b_0+b_D)m_\pi^2-(2C_1-C_7+b_4) zq^2+(2C_1-C_7+b_4+2C_2-C_8+b_8)w_\pi^2],\,0\Bigg\},
\end{align}
\begin{align}
\label{eq76}
T_{K \Sigma}^{(3/2,\text{NLO})}=&\Bigg\{\frac{1}{f_K^2}[-2(2b_0+b_D+b_F)m_K^2-2C_1 zq^2+2(C_1+C_2)w_K^2],\,-\frac{2}{f_K^2}C_3\Bigg\},
\end{align}
\begin{align}
\label{eq77}
T_{K \Sigma}^{(1/2,\text{NLO})}=&\Bigg\{\frac{1}{2f_K^2}[-4(2b_0+b_D-2b_F)m_K^2+(2C_1-3C_4+3C_7) zq^2-(2C_1-3C_4+3C_7\nonumber\\
&+2C_2-3C_5+3C_8)w_K^2],\,\frac{1}{4f_K^2}(4C_3+3C_6-3C_9)\Bigg\},
\end{align}
\begin{align}
\label{eq78}
T_{\overline{K} \Sigma}^{(3/2,\text{NLO})}=&\Bigg\{\frac{1}{2f_K^2}[-4(2b_0+b_D-b_F)m_K^2-2(C_4-C_7)zq^2+2(C_4-C_7\nonumber\\
&+C_5-C_8)w_K^2],\,-\frac{1}{2f_K^2}(C_6-C_9)\Bigg\},
\end{align}
\begin{align}
\label{eq79}
T_{\overline{K} \Sigma}^{(1/2,\text{NLO})}=&\Bigg\{\frac{1}{2f_K^2}[-4(2b_0+b_D+2b_F)m_K^2-(6C_1-C_4+C_7) zq^2+(6C_1-C_4+C_7\nonumber\\
&+6C_2-C_5+C_8)w_K^2],\frac{1}{4f_K^2}(12C_3+C_6-C_9)\Bigg\},
\end{align}
\begin{align}
\label{eq80}
T_{K\Xi}^{(1,\text{NLO})}=&\Bigg\{\frac{1}{f_K^2}[-2(2b_0+b_D+b_F)m_K^2-2C_1zq^2+2(C_1+C_2)w_K^2],\,\frac{2}{f_K^2}C_3\Bigg\},
\end{align}
\begin{align}
\label{eq81}
T_{K\Xi}^{(0,\text{NLO})}=&\Bigg\{\frac{2}{f_K^2}[(-2b_0-3b_D+b_F)m_K^2+(C_1-2C_4)zq^2-(C_1-2C_4\nonumber\\
&+C_2-2C_5)w_K^2],\frac{2}{f_K^2}(-C_3+C_6)\Bigg\},
\end{align}
\begin{align}
\label{eq82}
T_{\overline{K}\Xi}^{(1,\text{NLO})}=&\Bigg\{\frac{1}{f_K^2}[-4(b_0+b_D)m_K^2-2C_4zq^2+2(C_4+C_5)w_K^2],\,-\frac{1}{f_K^2}C_6\Bigg\},
\end{align}
\begin{align}
\label{eq83}
T_{\overline{K}\Xi}^{(0,\text{NLO})}=&\Bigg\{-\frac{1}{f_K^2}[4(b_0+b_F)m_K^2
+2(2C_1-C_4)zq^2-2(2C_1-C_4+2C_2-C_5)w_K^2],\nonumber\\
&\frac{1}{f_K^2}(-4C_3+C_6)\Bigg\},
\end{align}
\begin{align}
\label{eq84}
T_{K\Lambda}^{(\text{NLO})}=&\Bigg\{\frac{1}{6f_K^2}[-4(6b_0+5b_D)m_K^2-(-8b_4+2C_1+9C_4-C_7)zq^2+(-8b_4+2C_1+9C_4-C_7\nonumber\\
&-8b_8+2C_2+9C_5-C_8)w_K^2],\frac{1}{4f_K^2}(4C_3-C_6+C_9)\Bigg\},
\end{align}
\begin{align}
\label{eq85}
T_{\overline{K}\Lambda}^{(\text{NLO})}=&\Bigg\{\frac{1}{6f_K^2}[-4(6b_0+5b_D)m_K^2-(-8b_4+2C_1+9C_4-C_7)zq^2+(-8b_4+2C_1+9C_4-C_7\nonumber\\
&-8b_8+2C_2+9C_5-C_8)w_K^2],\,\frac{1}{4f_K^2}(-4C_3+C_6-C_9)\Bigg\},
\end{align}
\begin{align}
\label{eq86}
T_{\eta N}^{(\text{NLO})}=&\Bigg\{\frac{1}{3f_\eta^2}[-16(b_0+b_D-b_F)m_K^2+2(2b_0+3b_D-5b_F)m_\pi^2+2(2b_4+C_1-3C_4+C_7)zq^2\nonumber\\
&-2(2b_4+C_1-3C_4+C_7+2b_8+C_2-3C_5+C_8)w_\eta^2],\,0\Bigg\},
\end{align}
\begin{align}
\label{eq87}
T_{\eta \Sigma}^{(\text{NLO})}=&\Bigg\{\frac{2}{3f_\eta^2}[-8b_0 m_K^2+2(b_0-b_D)m_\pi^2-(b_4+2C_1-C_7)zq^2+(b_4+2C_1-C_7+b_8\nonumber\\
&+2C_2-C_8)w_\eta^2],\,0\Bigg\},
\end{align}
\begin{align}
\label{eq88}
T_{\eta \Xi}^{(\text{NLO})}=&\Bigg\{\frac{1}{3f_\eta^2}[-16(b_0+b_D+b_F)m_K^2+2(2b_0+3b_D
+5b_F)m_\pi^2-(-4b_4+4C_1+3C_4+C_7)zq^2\nonumber\\
&+(-4b_4+4C_1+3C_4+C_7-4b_8+4C_2+3C_5+C_8)w_\eta^2],\,0\Bigg\},
\end{align}
\begin{align}
\label{eq89}
T_{\eta \Lambda}^{(\text{NLO})}=&\Bigg\{\frac{2}{9f_\eta^2}[-8(3b_0+4b_D)m_K^2+2(3b_0+7b_D)m_\pi^2-9(b_4+C_4)zq^2+9(b_4+C_4+b_8\nonumber\\
&+C_5)w_\eta^2],\,0\Bigg\}.
\end{align}

Here we have introduced the nine linear combinations:
\begin{align}
\label{eq90}
&C_{1}=b_{1}+b_{2}+2b_{3},\quad
C_{2}=b_{5}+b_{6}+2b_{7},\quad
C_{3}=b_{9}+b_{10},\nonumber\\
&C_{4}=2b_{1}+2b_{3}+b_{4},\quad
C_{5}=2b_{5}+2b_{7}+b_{8},\quad
C_{6}=4b_{10}+b_{11},\nonumber\\
&C_{7}=2b_{2}-2b_{3}+b_{4},\quad
C_{8}=2b_{6}-2b_{7}+b_{8},\quad
C_{9}=4b_{9}+b_{11},
\end{align}
of the low-energy constants $b_{i}(i=1, ... , 11)$ in order to get a more compact representation.
\begin{figure}[t]
\centering
\includegraphics[height=12cm,width=8cm]{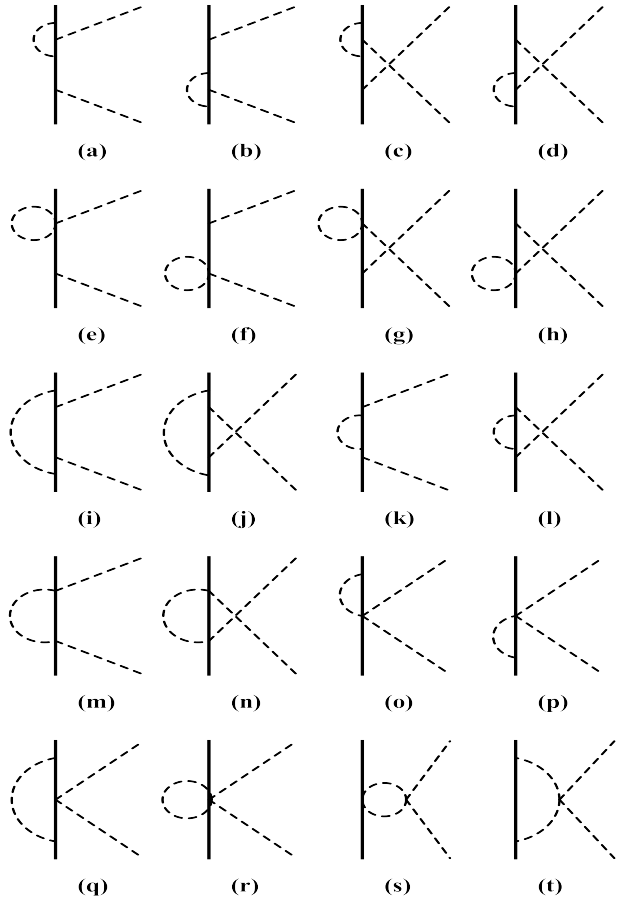}
\caption{\label{fig:knloopfeynman}Nonvanishing one-loop diagrams contributing at third chiral order. Diagrams with self-energy correction on external meson or baryon lines are not shown.  } 
\end{figure}

\subsection{Next-to-next-to-leading order amplitudes}
At third order $\mathcal{O}(q^{3})$ one has contributions from one-loop diagrams and counterterms. The nonvanishing one-loop diagrams generated by the vertices of $\mathcal{L}_{\phi\phi}^{(2)}$ and $\mathcal{L}_{\phi B}^{(1)}$ are shown in Fig.~\ref{fig:knloopfeynman}. The counterterm contributions estimated from resonance exchange were found to be much smaller than the chiral loop contributions in the case of the $\pi N$ scattering lengths \cite{bern1993,bern1995}. This feature may, however, not apply to meson-baryon scattering in general. In order to make our calculations involve fewer constants and to give a suitable comparison with the results of covariant baryon chiral perturbation theory, we are not considering the counterterm contributions when calculating T-matrices for most channels at $\mathcal{O}(q^{3})$ in this paper. But for improving the fits of $\pi N$ scattering, we are considering the counterterm contributions to the $\pi N$ channels at $\mathcal{O}(q^{3})$. The pertinent amplitudes can be obtained from the counterterm contributions in SU(2) HB$\chi$PT \cite{fett1998}. After redefining the LECs as $h_{1,2,3,4}$, the $\pi N$ amplitudes for total isospin $I=3/2$ and $I=1/2$ read:
\begin{align}
\label{eq91}
T_{\pi N}^{(3/2,\text{N2LO})}=&\Bigg\{\frac{w_\pi}{f_\pi^2}(-h_1 m_\pi^2+h_2 t-h_3 w_\pi^2),\,\frac{w_\pi}{f_\pi^2}h_4 \Bigg\},
\end{align}
\begin{align}
\label{eq92}
T_{\pi N}^{(1/2,\text{N2LO})}=&\Bigg\{\frac{2 w_\pi}{f_\pi^2}(h_1 m_\pi^2-h_2 t+h_3 w_\pi^2),\,\frac{w_\pi}{f_\pi^2}h_4\Bigg\}
\end{align}
with $t=2q^2(z-1)$ the invariant momentum transfer squared. The one-loop amplitudes are tedious to evaluate, and we give only the occurring loop functions in the Appendix. We use dimensional regularization and minimal subtraction to evaluate divergent loop integrals \cite{hoof1979,bern19951,mojz1998,bouz2000,bouz2002}. We use $f_{\pi}$, $f_{K}$, $f_\eta$ for pion, kaon, eta-baryon scattering amplitudes from loops instead of the value $f$ in the chiral limit. The differences only appear at higher order. Moreover, we have performed the renormalization of the baryon and meson masses and the meson-baryon coupling constants. In that procedure loop contributions proportional to $1/w$ from diagrams (a-h), (k) and (l) are subtracted and get subsumed in the leading order amplitudes written in terms of the physical parameters. The differences between $f$ and $f_{\pi, K, \eta}$ in the leading order amplitudes are balanced in the same way by loop contributions.

\section{Calculating phase shifts and scattering lengths}
\label{phase}
The partial wave amplitudes $f_{j}^{(I)}(q)$, where $j=l\pm 1/2$ refers to the total angular momentum and $l$ to orbital angular momentum, are obtained from the non-spin-flip and spin-flip amplitudes by a projection:
\begin{align}
\label{eq93}
f_{l\pm 1/2}^{(I)}(q)=\frac{M_{B}}{8\pi(w_{M}+E_{B})}\int_{-1}^{+1}dz\Big\{V_{M B}^{(I)}(q)P_{l}(z)+q^{2}W_{M B}^{(I)}(q)[P_{l\pm 1}(z)-zP_{l}(z)]\Big\},
\end{align}
where $P_{l}(z)$ denotes the conventional Legendre polynomial and $w_{M}+E_{B}=\sqrt{m_M^2+q^2}+\sqrt{M_B^2+q^2}$ is the total center-of-mass energy. For the energy range considered in this paper, the phase shifts $\delta_{l\pm 1/2}^{(I)}(q)$ are calculated by (also see Refs.~\cite{gass1991,fett1998})
\begin{align}
\label{eq94}
\delta_{l\pm 1/2}^{(I)}(q)=\text{arctan}[q\,\text{Re}\,f_{l\pm 1/2}^{(I)}(q)].
\end{align}
Based upon relativistic kinematics, there is the relation  between the center-of-mass momentum $q$ and the momentum $p_{\text{lab}}$ of the incident meson in the laboratory system, 
\begin{align}
\label{eq95}
&q^{2}=\frac{M_{B}^{2}p_{\text{lab}}^{2}}{m_{M}^{2}+M_{B}^{2}+2M_{B}\sqrt{m_{M}^{2}+p_{\text{lab}}^{2}}}.
\end{align}

The scattering lengths for s-waves and the scattering volumes for p-waves are obtained by approaching the threshold \cite{eric1988}
\begin{align}
\label{eq96}
a_{l\pm 1/2}^{(I)}=\lim\limits_{q \rightarrow 0}q^{-2l}f_{l\pm 1/2}^{(I)}(q).
\end{align}

\section{Results and discussion}
\label{results}
Before making predictions, we have to determine the low-energy constants. The parameters $M_0,b_D,b_F$ and $b_0$ have been determined in our previous paper \cite{huan2015}. In this paper, we take the same value $M_0=646\,$MeV, but the parameters $b_0,b_D$ and $b_F$ are obtained by fitting the phase shifts of $\pi N$ and $K N$ scattering. Here we introduce three linear combinations:
\begin{align}
\label{eq97}
B_1=2b_0+b_D+b_F,\quad
B_2=b_0+b_D,\quad
B_3=b_0-b_F.
\end{align}
After the regrouping in Eq.~(90), we also need to determine the nine combinations $C_{1, ... ,9}$. Throughout this paper, we use $m_{\pi}=139.57$ MeV, $m_{K}=493.68$ MeV, $m_{\eta}=547.86$ MeV, $M_{N}=938.9$ MeV, $M_{\Sigma}=1193.2$ MeV, $M_{\Xi}=1318.3$ MeV,  $M_{\Lambda}=1115.7$ MeV, $f_{\pi}=92.07$ MeV, $f_{K}=110.03$ MeV, and $f_{\eta}=1.2f_{\pi}$ \cite{pdg2016}. We also set the scale $\lambda$ equal to the chiral symmetry breaking scale, $\lambda=4\pi f_{\pi}=1.16\,\text{GeV}$. For the axial vector coupling constants we take the $D=0.8$ and $F=0.5$ as their physical values.

\begin{figure}
\centering
\includegraphics[height=12cm,width=12cm]{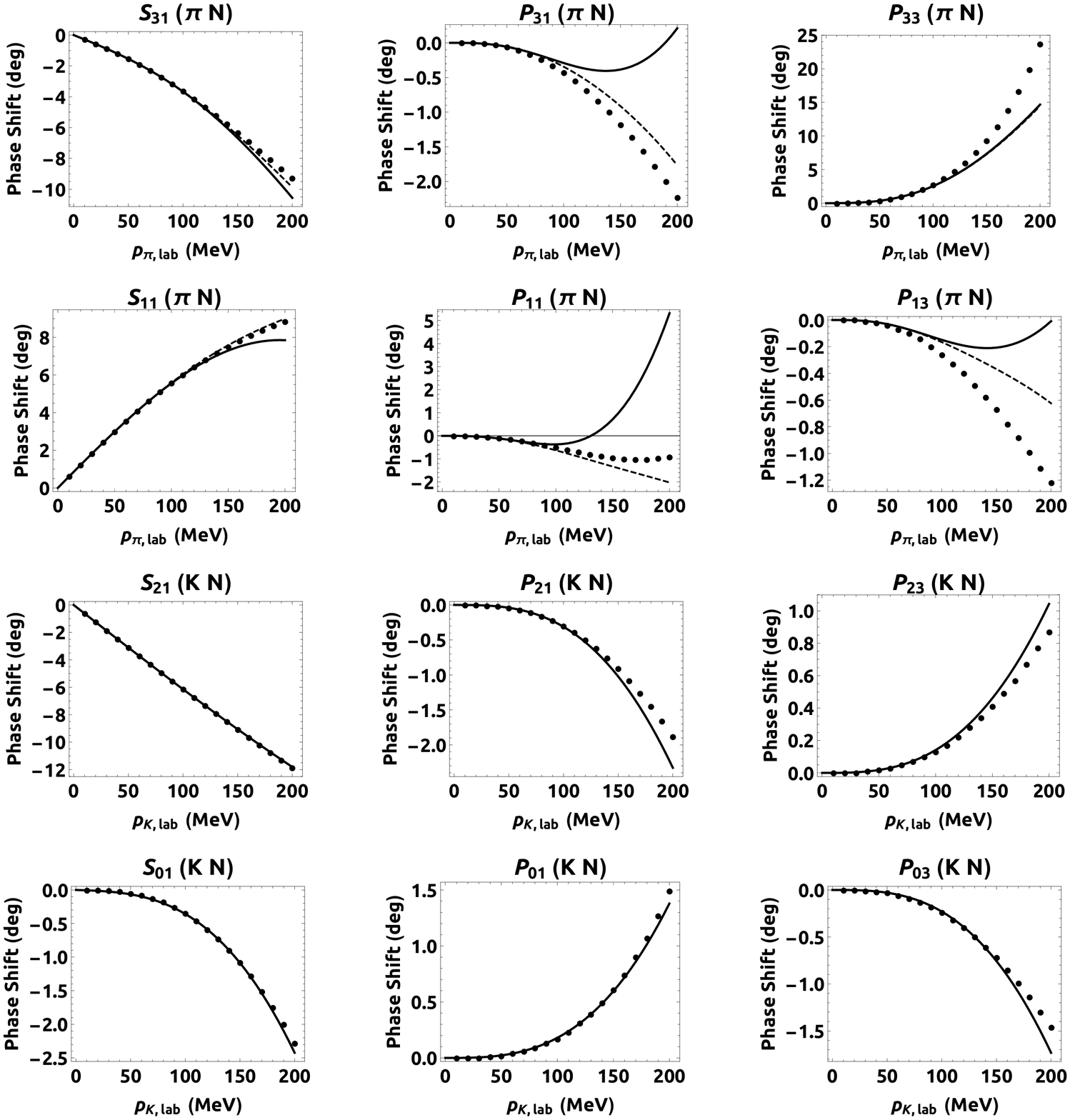}
\caption{\label{fig:Fitone} Fits and predictions for the WI08 and SP92 phase shifts in $\pi N$ and $KN$ scatterings versus the pion and kaon laboratory momenta, respectively. Fitting in all $\pi N$ waves except for $P_{11}$ wave and $KN$ waves of isospin $I=1$ are the data between 50 and 90 MeV, while fitting $P_{11}$ wave data from 30 to 70 MeV and $K N$ waves of isospin $I=0$ data at $p_\text{lab}=(110,120,130,140,150)$ MeV. For higher and lower energies, the phase shifts are predicted. The dotted lines denote the results from SAID \citep{SAID}. The solid lines refer to our calculations in SU(3) HB$\chi$PT and the dashed lines denote the results in SU(2) HB$\chi$PT for $\pi N$ scattering.} 
\end{figure}

We determine the 16 LEC combinations $B_{1,2,3}$, $C_{1, ..., 9}$ and $h_{1,...,4}$ by using the phase shifts of the WI08 solution and the SP92 solution for $\pi N$ and $K N$ scattering \cite{work2012,hysl1992,SAID}, respectively. Since both solutions include no uncertainties for the phase shifts, we choose a common uncertainty of $\pm4\%$ to all phase shifts except for $P_{11}$-wave of $\pi N$ scattering before the fitting procedure. The phase shifts in the $P_{11}$-wave of $\pi N$ scattering have a large errors, which e.g. amounts to $35\%$ at $77.35\,$MeV pion lab-momentum \cite{SAID}. Therefore, we give an uncertainty of $\pm20\%$ to the phase shifts of $P_{11}$-wave for $\pi N$ scattering. Actually, this uncertainty is arbitrary and it does not change the quality of the fits to phase shifts but only the respective $\chi^2/\text{dof}$. For fitting the parameters $B_1$, $C_{1,2,3}$ and $h_{1,2,3,4}$, we use the six $S$- and $P$-wave phase shifts for $\pi N$ scattering between 50 and 90 MeV pion lab-momentum. For $P_{11}$-wave, we use the phase shifts between 30 and 70 MeV pion lab-momentum to make the fitting better. In addition, we can determine the parameters $B_1$, $C_{1,2,3}$ and $h_{1,2,3,4}$ in both SU(2) and SU(3) HB$\chi$PT. The difference comes from kaon and eta-loop contributions. The one-loop contributions in SU(2) HB$\chi$PT can be found in Ref.~\cite{bern1997}. We obtain in SU(3) HB$\chi$PT
\begin{align}
\label{98}
B_1^{(3)}=-2.80\pm 0.55,\quad
C_1^{(3)}=-8.29\pm 0.12,\quad
C_2^{(3)}=6.55\pm 0.48,\quad
C_3^{(3)}=2.82\pm 0.05,\nonumber\\
h_1^{(3)}=12.63\pm 5.61,\quad
h_2^{(3)}=5.99\pm 0.45,\quad
h_3^{(3)}=-15.46\pm 4.71,\quad
h_4^{(3)}=-22.51\pm 0.90
\end{align}
with a $\chi^2/\text{dof}\simeq 1.0$ and in SU(2) HB$\chi$PT
\begin{align}
\label{99}
B_1^{(2)}=-2.74\pm 0.34,\quad
C_1^{(2)}=-6.63\pm 0.07,\quad
C_2^{(2)}=4.09\pm 0.29,\quad
C_3^{(2)}=2.05\pm 0.03,\nonumber\\
h_1^{(2)}=8.00\pm 3.41,\quad
h_2^{(2)}=5.71\pm 0.27,\quad
h_3^{(2)}=-10.54\pm 2.86,\quad
h_4^{(2)}=-8.37\pm 0.55
\end{align}
with a $\chi^2/\text{dof}\simeq 0.4$. Note that, throughout this paper, the parameters $B_{1,2,3}$ and $C_{1,...,9}$ are in units of $\text{GeV}^{-1}$ while $h_{1,...,4}$ are in units of $\text{GeV}^{-2}$. The partial waves are denoted by $L_{2I,2J}$ with $L$ the angular momentum, $I$ the total isospin, and $J$ the total angular momentum. If $I$ is fixed, we use the simpler notation $L_{2J}$. For fitting $B_2$ and $C_{4,5,6}$ we use the data in the $S_{21}$, $P_{21}$ and $P_{23}$ partial waves of KN scattering between 50 and 90 MeV kaon lab-momentum. For determining $B_3$ and $C_{7,8,9}$, we fit to the data of the $S_{01}$, $P_{01}$, and $P_{03}$ partial waves at $p_K=(110,120,130,140,150)$ MeV. The resulting LECs have the values
\begin{align}
\label{100}
B_2=-0.43\pm 1.91,\quad
C_4=-2.04\pm 0.05,\quad
C_5=2.35\pm 3.79,\quad
C_6=8.29\pm 0.08
\end{align}
with a $\chi^2/\text{dof}\simeq 0.9$ and 
\begin{align}
\label{101}
B_3=-0.97\pm 0.07, \quad
C_7=-1.41\pm 0.04,\quad
C_8=3.35\pm 0.13,\quad
C_9=4.95\pm 0.05
\end{align}
with a $\chi^2/\text{dof}\simeq 0.4$. The uncertainty for the respective parameters is purely statistical and it measures how much a particular parameter can be changed while maintaining a good description of the fitted data, as detailed in Refs.~\cite{doba2014,carl2016}. The corresponding $S$- and $P$-wave phase shifts are shown in Fig.~\ref{fig:Fitone}. For all $S$-waves, we can obtain a good reproduction. Unfortunately, we failed to predict the $P$-wave phase shifts above $p_{\pi,\text{lab}}=100$ MeV for $\pi N$ scattering in SU(3) HB$\chi$PT. Interestingly, it is better in SU(2) HB$\chi$PT. The reason for this are large effects from loop diagrams involving internal kaon lines whose contributions to the $\pi N$ scattering amplitudes cannot be absorbed in the counterterms. For the partial waves in Fig.~\ref{fig:Fitone}, however, the description of the phase shifts is in agreement with the empirical phase shifts below 100 MeV.

Now we have fixed the 16 LEC combinations $B_{1,2,3}$, $C_{1, ..., 9}$ and $h_{1,...,4}$. But there are still two more parameters, namely $b_{4}$ and $b_{8}$. We have no further data available   to fix these two parameters. Therefore, we use the values determined from elastic and inelastic scatterings \cite{ramo2000} using coupled channels. Comparing our heavy-baryon Lagrangian $\mathcal{L}_{\phi B}^{(2,\text{ct})}$ with the corresponding Lagrangian of Refs.~\cite{kais1997,ramo2000}, one obtains the relations $2(b_{4}+b_{8})=d_{1}$ and $-2b_{4}=g_{1}$. This gives us the estimates  $b_4=-0.73\,\,\text{GeV}^{-1}$ and $b_8=0.81\,\,\text{GeV}^{-1}$.

In order to check the consistency of the ChPT framework for different observables, we determine some combinations of the low-energy constants from scattering lengths. We evaluate T-matrices at threshold ($q=0$) to obtain the scattering lengths in SU(3) HB$\chi$PT. We use the four scattering lengths $a_{\pi N}^{(3/2)}\simeq-0.11$ fm and $a_{\pi N}^{(1/2)}\simeq0.24$ fm from the WI08 solution \cite{work2012,SAID}, $a_{KN}^{(1)}\simeq-0.33$ fm and  $a_{KN}^{(0)}\simeq0.00$ fm from the SP92 solution \cite{hysl1992} and determine the four LEC combinations as
\begin{align}
\label{102}
B_1-C_1-C_2\simeq-1.08,\quad
h_1+h_3\simeq-3.05,\quad
2B_2-C_4-C_5\simeq-1.46,\quad
2B_3+C_7+C_8\simeq0.00.
\end{align}
The same combinations of low-energy constants determined from fitting to phase shifts are $B_1-C_1-C_2\simeq-1.06$ [see Eq.~(98)], $h_1+h_3\simeq-2.83$ [see Eq.~(98)], $2B_2-C_4-C_5\simeq-1.17$ [see Eq.~(100)] and  $2B_3+C_7+C_8\simeq0.00$ (see Eq.~(101)). These results are within errors consistent with the LEC combinations in Eq.~(102). In addition, the combination $b_0+b_D+b_F\simeq-1.13$ also is within error consistent with the values $-1.46$ determined by baryon masses from Ref.~\cite{huan2015}.

\begin{figure}
\centering
\includegraphics[height=24cm,width=12cm]{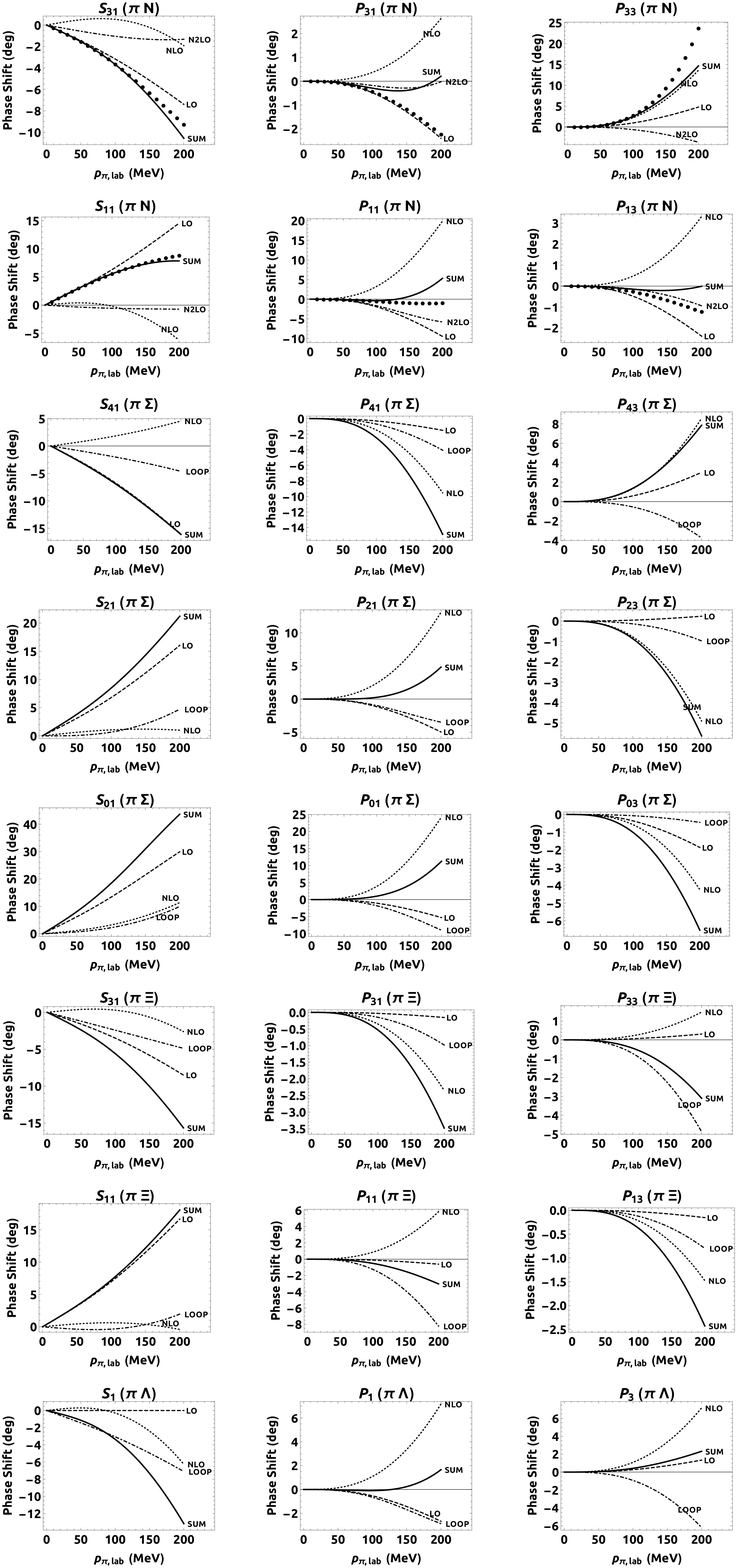}
\caption{\label{fig:PrePiBaryon} Predictions for the pion-baryon phase shifts versus the pion laboratory momentum. The dashed, dotted, dash-dotted  and solid lines denote the first-, second-, third-order and their sum contributions, respectively. The dotted lines denote the results from WI08 solution \citep{SAID}.} 
\end{figure}

\begin{figure}
\centering
\includegraphics[height=12cm,width=12cm]{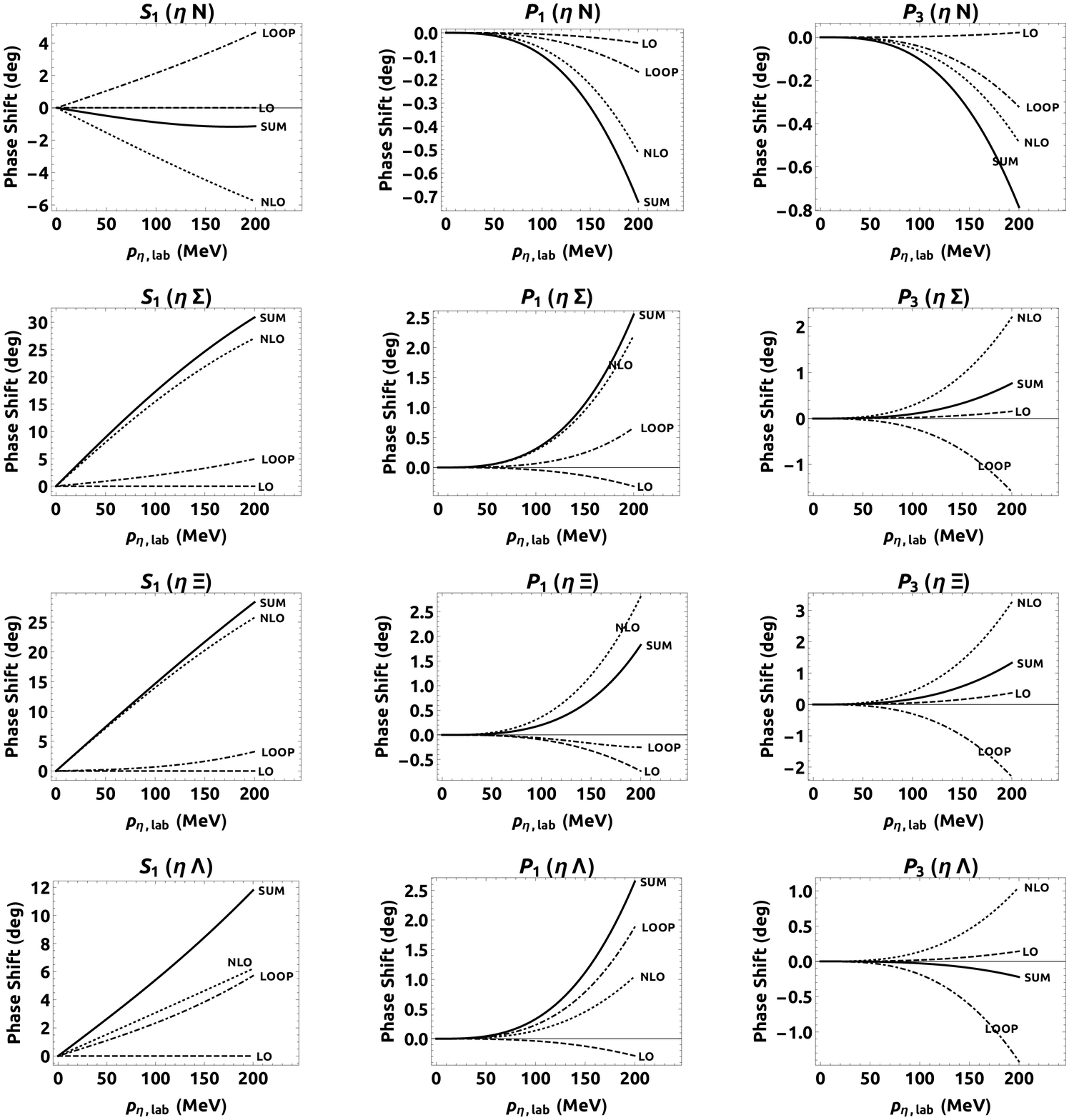}
\caption{\label{fig:PreEtaBaryon} Predictions for the eta-baryon phase shifts versus the eta laboratory momentum. The notation for lines is the same as in Fig.~\ref{fig:PrePiBaryon}.} 
\end{figure}

Next, we can predict phase shifts for octet-meson-baryon scattering which are shown in Figs.~\ref{fig:PrePiBaryon}\nobreakdash--\ref{fig:PreEtaBaryon}. First, let us look at $\pi N$ scattering since there exist empirical phase shifts for comparison. For all partial waves of $\pi N$ scattering, our results are consistent with empirical phase shifts except for $P$-waves at high energies $p_{\pi,\rm{lab}}>150\,$MeV. This feature originate from one-loop diagrams involving internal kaon lines. For the $P_{31}$-wave, the leading-order contribution is in good agreement with empirical phase shifts at the energies considered. This means that all higher order contributions should cancel out, to maintain the good description of the data. For the $P_{33}$-wave, the leading-order contribution is much smaller than the empirical phase shift below $200\,$MeV lab-momentum. In fact, there exists the prominent resonance $\Delta(1232)$ in this channel. One can also obtain a good description of the phase shifts at higher energies through introducing the $\Delta(1232)$ resonance explicitly in the effective Lagrangian \cite{elli1998}. For the $P_{11}$-wave, there also exists a resonance $N^*(1440)$, known as the Roper resonance. Nevertheless, Fettes \textit{et al.} \cite{fett1998} successfully predicted the phase shifts of the $P_{11}$-wave below $250\,$MeV in SU(2) HB$\chi$PT. But we must emphasize that the empirical phase shifts have large errors in this wave.

\begin{figure}
\centering
\includegraphics[height=6cm,width=12cm]{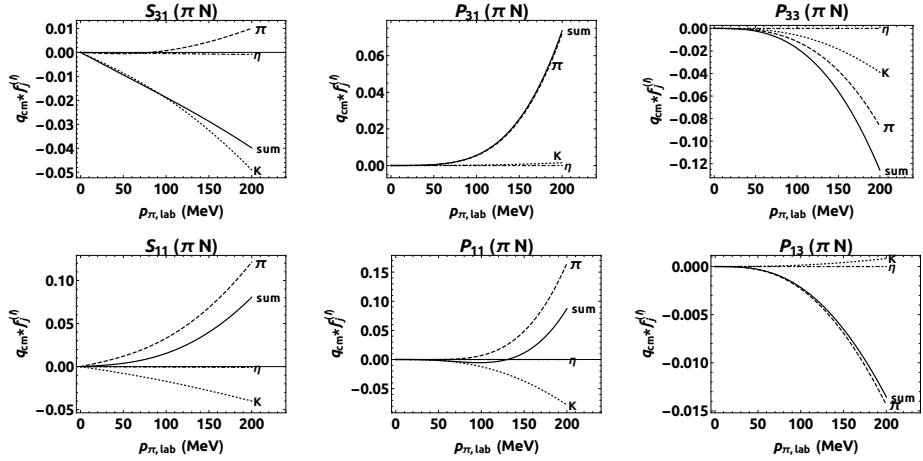}
\caption{\label{fig:PiNLOOPpiKeta}Contributions involving different mesons ($\pi, K, \eta$) internal lines  from one-loop diagrams at third chiral order are shown as the real part of amplitudes multiplied by the center-of-mass momentum. The dashed, dotted, dash-dotted and solid lines denote the contributions from mesons ($\pi, K, \eta$) internal lines and total one-loop diagrams, respectively.}
\end{figure}

Let us revisit $\pi N$ scattering in SU(2) HB$\chi$PT. The pertinent formulas can be found in Refs.~\cite{fett1998,bern1997}. We succeed to fit the six partial waves of $\pi N$ scattering only if the third-order counterterms are included. But we are able to fit the phase shifts of $S_{31}$-, $P_{31}$- and $P_{33}$-waves below 100 MeV pion lab-momentum without the third-order counterterms. The situation is the same in SU(3) HB$\chi$PT. This means it is important to consider the third-order counterterms for fitting the six partial waves. We can make a direct comparison between the third-order counterterms and the one-loop corrections through fitting the six partial waves of $\pi N$ scattering. We consider three different scale parameters introduced by dimensional regularization $\lambda=m_\pi,(m_\pi+4\pi f_\pi)/2,4\pi f_\pi$. The choice $\lambda=m_\pi$ makes the chiral logarithms $\text{ln}(m_\pi/\lambda)$ disappear, which leads one the scale-independent (barred) LECs in Ref.~\cite{fett1998}. For S-waves, we find that the third-order counterterm contribution is much smaller than the one-loop correction. This is consistent with results where the third-order counterterm contributions were estimated from resonance exchange \cite{bern1993,bern1995}. But for P-waves, the situation is more complicated. For the $P_{31}$-wave, the third-order counterterm contribution is always larger than the one-loop correction at high energies. For the $P_{11}$- and $P_{13}$-wave, we still obtain the same results except for the case of $\lambda=m_\pi$. For the $P_{33}$-wave, the counterterm contribution always almost cancels the one-loop correction. One needs to calculate the phase shifts at higher order for obtaining an improved result, as done in Ref.~\cite{fett2000}. We check that the leading-order amplitudes from SU(2) HB$\chi$PT are same as those from SU(3) HB$\chi$PT if we set axial vector coupling constants $g_A=D+F$. In fact, the amplitudes from the tree diagram in SU(2) and SU(3) HB$\chi$PT should be consistent because of conservation of strangeness. The difference comes from loop diagrams because there exist kaon and eta in meson internal lines. We make a direct comparison about contributions involving different mesons ($\pi, K, \eta$) internal lines from one-loop diagrams at third chiral order, see Fig.~\ref{fig:PiNLOOPpiKeta}. We can find that the contributions involving kaon internal lines are not small in $S_{31}$-, $P_{33}$-, $S_{11}$-, and $P_{11}$-waves. The difference of the amplitudes from one-loop diagrams between SU(2) and SU(3) HB$\chi$PT has no effect on the fitting at low energies. But the large contributions involving kaon internal lines make some effect on the fitting for $P$-waves at high energies. However, the third-order counterterms with LECs are also important to achieve a good fitting in covariant baryon chiral perturbation theory \cite{alar2013}. To sum up, for fitting the six partial waves of $\pi N$ scattering at third order, the third-order counterterm contributions should be considered. Fortunately, the third-order counterterms are the same in SU(2) and SU(3) HB$\chi$PT because of conservation of strangeness. Meanwhile, the minimal third-order meson-baryon Lagrangian consisting of 78 independent operators has been constructed in Refs.~\cite{olle2006,frin2006}. Moreover, the fourth-order meson-baryon chiral Lagrangian has also been constructed, see Ref.~\cite{jian2017}. In addition, the inclusion of the $\Delta$ resonance substantially improves the convergence for $\pi N$ scattering \cite{siem2016,blin2014}. The calculations including these counterterm contributions in other channels and the $\Delta$ resonance in $\pi N$ channel  will be done in forthcoming works.

Second, we predict directly the phase shifts of $\pi \Lambda$ scattering at the $\Xi$ mass ($p_{\pi,\text{lab}}\simeq 160$ MeV). For the $j=1/2$ channels, the predicted phase shifts
\begin{align}
\label{103}
\delta_{S}=(-8.37^{+0.06}_{-0.06})^{\circ},\quad\delta_{P}=(0.41^{+0.18}_{-0.21})^{\circ},\quad\delta_P-\delta_S=(8.78^{+0.19}_{-0.22})^{\circ}
\end{align}
are in agreement with the experiment results from FNAL E756 \cite{chak2003} [$D_{Str}+d_{CP}=(3.17\pm5.28\pm0.73)^{\circ}$] and HyperCP \cite{huan2004} [$D_{Str}+d_{CP}=(4.6\pm1.4\pm1.2)^{\circ}$] within errors. The uncertainties are from the uncertainties of the LECs through propagation of uncertainty  and the variation of the renormalization scale $\lambda$ ($0.94\, \text{GeV}<\lambda<1.31 \,\text{GeV}$). Of course, the parameters $b_4$ and $b_8$ are not determined by our formula. We just estimate them from the results employing the coupled channel approach. If we allow them to vary between zero and twice their values, the phase shifts do almost not change. That means the phase shifts are not sensitive to these two parameters. Note that, in the  hyperon decay $\Xi^{-}\rightarrow\Lambda\pi^-$, $D_{Str}=\delta_P-\delta_S$ is the strong rescattering phase shift difference that is given by the phase shift difference of $\pi \Lambda$ elastic scattering at the $\Xi$ mass between the $P$- and $S$-wave, while $d_{CP}$ is weak CP-violating phase shift difference. The CP violation in this hyperon decay may be suppressed by a vanishing $d_{CP}$. It is consistent with the results from Refs.~\cite{holm2004,abda2006} that do not support CP-violation in this decay.

Third, we find that the phase shifts reach over 40 degrees at laboratory momenta of 200 MeV in the $S_{01}$($\pi \Sigma$), $S_{01}$($K \Xi$), $S_{01}$($\overline{K} N$), and $S_{11}$($\overline{K} \Sigma$) waves. It is possible to generate dynamically resonances in these waves because there exists strong enough attraction to support bound states or quasibound states. In fact, for strangeness $S=-1$ including $S_{01}$($\pi \Sigma$), $S_{01}$($K \Xi$), $S_{01}$($\overline{K} N$) waves, nonperturbative methods \cite{kais1995,oset1998,olle2001,lutz2002,oset2002,hyod2012} were used to generate dynamically resonances. $\Lambda(1405)$ is generated in the $S_{01}$($\pi \Sigma$) and $S_{01}$($\overline{K} N$) channels, while the attraction in the $S_{01}$($K \Xi$) channel is responsible for the $\Lambda(1670)$ resonance. It is natural to expect that a resonance is also generated in the $S_{11}$($\overline{K} \Sigma$) channel with strangeness $S=-2$. We find that the $S_{11}$($\overline{K} \Sigma$) threshold energy ($\sim$1687 MeV) is in good agreement with the mass of a three-star resonance $\Xi(1690)$ with strangeness $S=-2$ from PDG \cite{pdg2016}. Unfortunately, the experiment can not distinguish between a resonance and a large scattering length \cite{dion1978}. We suggest that the meson-baryon scattering for strangeness $S=-2$ can be studied in the chiral unitary approach based on coupled channels to generate dynamically resonance.
 
Fourth, it is interesting to study eta-baryon scattering. The $\eta$-mesic nuclei was predicted by Haider and Liu \cite{haid1986} using a $\eta N$ interaction model developed by Bhalerao and Liu \cite{bhal1985} over thirty years ago; for more detail about $\eta$-mesic nuclei, see the review paper \cite{haid2015}. The $\eta$-mesic hypernuclei were also proposed by Abaev and Nefkens \cite{abae1996} based on their predicted $\eta \Lambda$ scattering length. Indeed, we obtain a small repulsion in $S$-wave of $\eta N$ channel and attraction in $S$-wave of $\eta \Lambda$ channel. However, the phase shift does not reach over 40 degrees at 200 MeV, which the $S_{01}$($\pi \Sigma$), $S_{01}$($K \Xi$), $S_{01}$($\overline{K} N$), and $S_{11}$($\overline{K} \Sigma$) waves corresponding to respective resonances have the feature of the phase shift. Thus, the attraction presumably is not strong enough for the formation of $\eta$-mesic hypernuclei.

Then we calculate the scattering lengths through the use of Eq.~(96) order by order. The results are shown in Table~\ref{tab:MesonBaryonScatteringLengths}.
The scattering lengths are obtained by using an incident meson momentum $p_{\text{lab}}=10$ MeV to approximate its value at threshold. The errors of the scattering lengths in our calculations are estimated from the statistical errors of the constants ($M_0$, $b_{0,D,F}$ and $C_{1,...,9}$) with the error propagation formula. There are not yet enough empirical data for comparison. Instead, we present the scattering lengths calculated through the threshold T-matrices from Refs.~\cite{liu20071,mai2009} in Table~\ref{tab:MesonBaryonScatteringLengths}. 
Our values except for $\eta$-baryon channels are in agreement with the results calculated by threshold T-matrices in the framework of HB$\chi$PT in Refs.~\cite{liu20071,mai2009} within errors. The results calculated in HB$\chi$PT from Ref.~\cite{mai2009} have larger errors because they considered the origins of the errors from both the uncertainties of experimental input and the variation of the renormalization scale $\lambda$ ($0.94\, \text{GeV}<\lambda<1.31 \,\text{GeV}$). Note that we only give the real parts of the scattering lengths in our paper. For the imaginary parts of the scattering lengths we obtain the almost same results as Ref.~\cite{liu20071}. For the $\eta N$ channel, we obtain negative scattering length the same as the result was calculated with decuplet contributions in Ref.~\cite{liu20072}. Thus, the repulsion indicates whether $\eta$-mesic nuclei are possible requires further investigations. We can also see that the large scattering lengths which reach over 1 fm were obtained in $\text{Re}\,a_{K\Xi}^{(0)}$, $\text{Re}\,a_{\overline{K}N}^{(0)}$ and $\text{Re}\,a_{\overline{K}\Sigma}^{(1/2)}$ channels with HB$\chi$PT. That is consistent with the prediction of the phase shift as mentioned above. For the $a_{\pi \Sigma}^{(0)}$ channel, the scattering length does not reach over 1 fm, but it is still larger than the other pion-baryon channel. Compared with the large phase shift, the scattering length is small. It is because the resonance $\Lambda(1405)$ is generated at $p_{\pi,\text{lab}}\simeq170$ MeV that is far away from threshold in this channel. Let us look at the scattering lengths calculated in the framework of infrared regularization (IR) from Ref.~\cite{mai2009}. First, for $\pi N$ scattering lengths, their predictions are not consistent with the experiment values ($a_{\pi N}^{(3/2)}=-0.125^{+0.003}_{-0.003}$ fm, $a_{\pi N}^{(1/2)}=0.250^{+0.006}_{-0.004}$ fm) from Ref.~\cite{schr2001}. Thus, their prediction for the small value of the $\pi \Lambda$ scattering length is questionable. Instead, our prediction in HB$\chi$PT for this channel is more reasonable; see the above statement for the phase-shift difference. Second, the convergence in IR is not improved. In contrast, for pion-baryon scatterings the scattering lengths from HB$\chi$PT show better convergence than those from IR. However, in both approaches the higher order contributions need to be calculated. 

Finally, we can check the convergence from both the phase shifts and scattering lengths. For $\pi N$ channels, the phase shifts in the $S_{31}$- and $S_{11}$-wave below $p_{\pi,\text{lab}}=100$ MeV and $P_{31}$-wave at energies considered and also the two scattering lengths are dominated by the leading-order contributions. That means the second-order and the third-order contributions should be cancelled out in any perturbative calculations up to third order. Thus, for checking the convergence in $\pi N$ channels, it is not enough to calculate the amplitudes at third order. Furthermore, some of the channels are also affected by the resonances such as $\Delta(1232)$ in the $P_{33}(\pi N)$ channel. Thus, it is not a surprise that we cannot obtain a reasonable convergence. However, we can also find some interesting features from the phase shifts and scattering lengths. For $S_{11}(\pi N)$, $S_{21}(\pi \Sigma)$ and $S_{11}(\pi \Xi)$ channels, they have the similar phase-shift feature. The phase shifts are dominated by the leading-order contributions. The contributions from next-to-leading-order and one-loop order amplitudes are small up to $p_{\pi,\text{lab}}\simeq100$ MeV. For $KN$ channels, the phase shifts are not dominated by leading-order contributions. In $S_{31}(K \Sigma)$, $S_{1}(K \Lambda)$,  $S_{21}(\overline{K}N)$, $S_1(\overline{K}\Lambda)$, $S_1(\eta \Sigma)$ and $S_1(\eta \Xi)$ channels, we find the one-loop contributions are much smaller than the next-to-leading-order contributions. Note that the six channels except for $S_{1}(K \Lambda)$ and $S_1(\overline{K}\Lambda)$ have imaginary parts in loop contributions. That may be the reason why the phase shifts from loop contributions are small. For the scattering lengths excluding the imaginary parts, the one-loop contributions are also smaller than the next-to-leading-order contributions. For P-waves, we find they have better convergence than S-waves. The convergence for meson-baryon scattering does not improve in the framework of infrared regularization, see Ref.~\cite{mai2009}. Thus, a higher-order calculation for meson-baryon scattering is needed.

In summary, we have calculated the T-matrices of meson-baryon scattering to one-loop order in SU(3) HB$\chi$PT. We fitted to phase shifts of $\pi N$ and $KN$ scattering in order to determine the LECs. Then we discussed the fitting in detail and pointed out that the counterterm contribution is important to fit $P$ waves. We predicted the other channels by using these LECs, and obtained the strong phase shift difference of $\pi \Lambda$ elastic scattering at the $\Xi$- mass between the P- and S-wave and the result is in fair agreement with the experiment result. We found that the phase shifts in $S_{01}$($\pi \Sigma$), $S_{01}$($K \Xi$), $S_{01}$($\overline{K} N$), and $S_{11}$($\overline{K} \Sigma$) waves are so strong that resonances may be generated dynamically in all these waves and suggest that the meson-baryon scattering for strangeness $S=-2$ can be calculated in chiral unitary approach. We also found the phase shifts in $\eta N$ scattering are small and negative. We calculated the scattering lengths order by order and made comparisons with the results derived from threshold T-matrices in both HB$\chi$PT and covariant formulations employing infrared regularizations (IR). We found the scattering lengths from IR do not show a better convergence of the chiral expansion for meson-baryon scattering. Finally, we discussed the convergence of the meson-baryon scattering. We expect to obtain improved results for meson-baryon scattering in future higher-order and nonperturbative calculations.

\section*{Acknowledgements}
This work is supported in part by the China Scholarship Council (CSC), the National Natural Science Foundation of China under Grants No. 11465021 and No. 11065010.  We thank Yan-Rui Liu (Shandong University) and Wolfram Weise (Technische Universit\"{a}t M\"{u}nchen) for very helpful discussions. This work is also supported in part by the DFG and the NSFC through funds provided by the Sino-German CRC 110.

\newpage

\begin{figure}
\centering
\includegraphics[height=21cm,width=12cm]{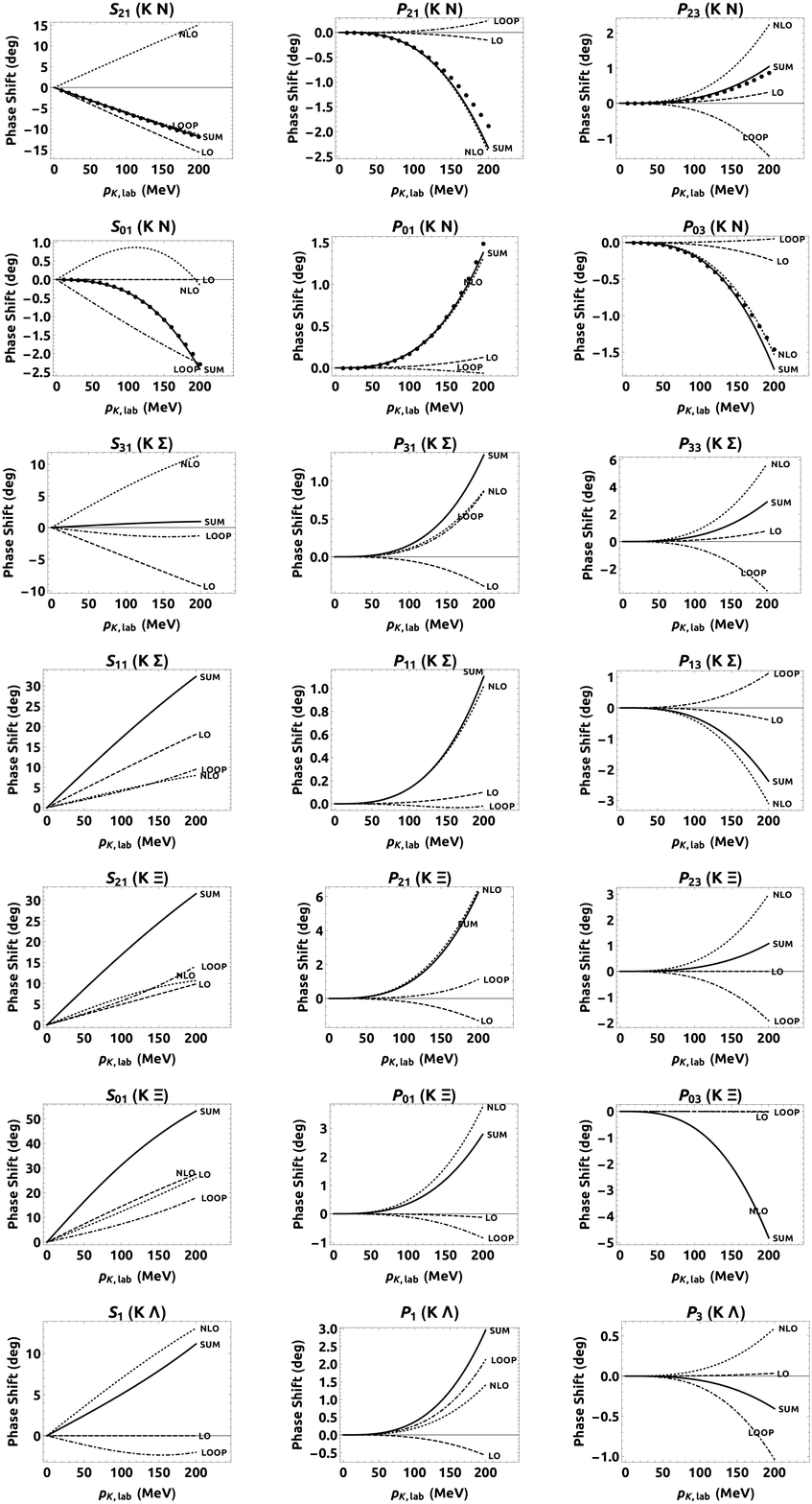}
\caption{\label{fig:PreKBaryon} Predictions for the kaon-baryon phase shifts versus the kaon laboratory momentum. The notation for lines is the same as in Fig.~\ref{fig:PrePiBaryon}. The dotted lines denote the results from SP92 solution \citep{SAID}.} 
\end{figure}

\begin{figure}
\centering
\includegraphics[height=21cm,width=12cm]{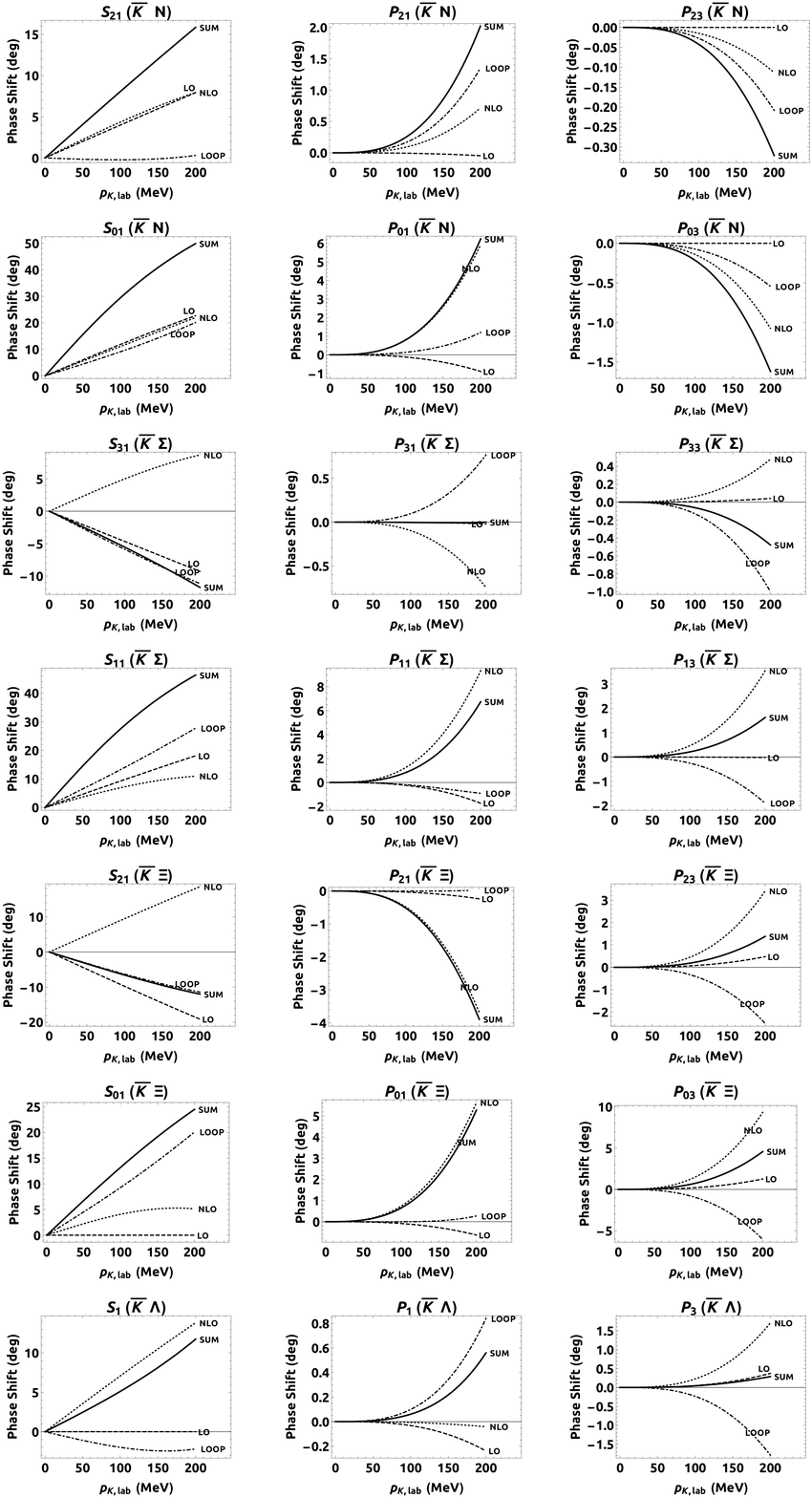}
\caption{\label{fig:PreKbarBaryon} Predictions for the antikaon-baryon phase shifts versus the antikaon laboratory momentum. The notation for lines is the same as in Fig.~\ref{fig:PrePiBaryon}.} 
\end{figure}

\begin{table*}
\centering
\begin{threeparttable}
\caption{\label{tab:MesonBaryonScatteringLengths}
Values of the meson-baryon scattering lengths for our calculations in comparison to the results from Refs.~\cite{liu20071,mai2009}. The scattering lengths are in units of fm.}
\begin{tabular}{ccccccccccc}
\midrule
\toprule
Channel & $\mathcal{O}(q)$ & $\mathcal{O}(q^2)$ & $\mathcal{O}(q^3)$ & Total & Liu(HB)\tnote{1} & Mai(HB)\tnote{2} & Mai(IR)\tnote{3} &\\
\midrule
$a_{\pi N}^{(3/2)}$&$-0.11$&$0.05^{+0.002}_{-0.002}$&$-0.05$&$-0.11^{+0.002}_{-0.002}$&$-0.130^{+0.001}_{-0.003}$
&$-0.13^{+0.03}_{-0.03}$&
$-0.04^{+0.07}_{-0.07}$&\\
$a_{\pi N}^{(1/2)}$&$0.23$&$0.05^{+0.002}_{-0.002}$&$-0.03$&$0.24^{+0.002}_{-0.002}$& $0.260^{+0.001}_{-0.003}$&$0.26^{+0.03}_{-0.03}$&
$0.07^{+0.07}_{-0.07}$&\\
\midrule
$a_{\pi \Sigma}^{(2)}$&$-0.23$&$0.06^{+0.001}_{-0.001}$&$-0.07$&$-0.24^{+0.001}_{-0.001}$& $-0.25$&$-0.24^{+0.01}_{-0.01}$ & $0.01^{+0.04}_{-0.04}$&\\
$a_{\pi \Sigma}^{(1)}$&$0.23$&$0.05^{+0.001}_{-0.001}$&$-0.00$&$0.28^{+0.001}_{-0.001}$& $0.26^{+0.03}_{-0.03}$&$0.33^{+0.06}_{-0.06} $ &
$0.10^{+0.16}_{-0.17}$ &\\
$a_{\pi \Sigma}^{(0)}$&$0.46$&$0.08^{+0.001}_{-0.001}$&$0.04$&$0.59^{+0.001}_{-0.001}$& $0.60^{+0.04}_{-0.04}$&$0.49^{+0.07}_{-0.08}$ &
$0.10^{+0.17}_{-0.19}$ &\\
\midrule
$a_{\pi \Xi}^{(3/2)}$&$-0.12$&$0.04^{+0.000}_{-0.000}$&$-0.09$&$-0.17^{+0.000}_{-0.000}$& $-0.17$& $-0.17^{+0.03}_{-0.03}$ & $0.02^{+0.06}_{-0.07}$&\\
$a_{\pi \Xi}^{(1/2)}$&$0.23$&$0.04^{+0.000}_{-0.000}$&$-0.03$&$0.24^{+0.000}_{-0.000}$& $0.23$& $0.23^{+0.03}_{-0.03}$ &$0.02^{+0.08}_{-0.10}$ &\\
\midrule
$a_{\pi \Lambda}$&$0.00$&$0.04^{+0.001}_{-0.001}$&$-0.11$&$-0.07^{+0.001}_{-0.001}$& $-0.071^{+0.004}_{-0.005}$& $-0.09^{+0.01}_{-0.01}$ &
$-0.01^{+0.04}_{-0.04}$ &\\
\midrule
$a_{K N}^{(1)}$&$-0.42$&$0.41^{+0.005}_{-0.005}$&$-0.32$&$-0.33^{+0.005}_{-0.005}$&$-0.33
$&$-0.33^{+0.10}_{-0.10}$ &$-0.33^{+0.32}_{-0.32}$ &\\
$a_{K N}^{(0)}$&$0.00$&$0.06^{+0.004}_{-0.004}$&$-0.07$&$-0.00^{+0.004}_{-0.004}$&$0.02
$&$0.02^{+0.27}_{-0.27}$ &$0.02^{+0.64}_{-0.64}$&\\
\midrule
$\text{Re}\, a_{K \Sigma}^{(3/2)}$&$-0.23$&$0.33^{+0.011}_{-0.011}$&$-0.07$&$0.03^{+0.011}_{-0.011}$
&$0.0024^{+0.0086}_{-0.0202}$&$-0.04^{+0.20}_{-0.19}$&$-0.28^{+0.52}_{-0.49}$&\\
$\text{Re}\, a_{K \Sigma}^{(1/2)}$&$0.45$&$0.22^{+0.005}_{-0.005}$&$0.19$&$0.86^{+0.005}_{-0.005}$
&$0.81^{+0.01}_{-0.01}$&$0.83^{+0.26}_{-0.30}$&$0.87^{+0.55}_{-0.64}$&\\
\midrule
$\text{Re}\,a_{K \Xi}^{(1)}$&$0.23$&$0.34^{+0.011}_{-0.011}$&$0.25$&$0.82^{+0.011}_{-0.011}$
&$0.82^{+0.01}_{-0.02}$&$0.72^{+0.19}_{-0.21}$&$0.48^{+0.43}_{-0.43}$&\\
$\text{Re}\,a_{K \Xi}^{(0)}$&$0.70$&$0.57^{+0.001}_{-0.001}$&$0.32$&$1.59^{+0.001}_{-0.001}$
&$1.46^{+0.01}_{-0.02}$&$1.48^{+0.19}_{-0.21}$&$1.02^{+0.51}_{-0.68}$&\\
\midrule
$a_{K \Lambda}$&$0.00$&$0.35^{+0.006}_{-0.006}$&$-0.11$&$0.24^{+0.006}_{-0.006}$
&$0.17^{+0.06}_{-0.06}$&$0.34^{+0.18}_{-0.19}$&$0.19^{+0.55}_{-0.56}$&\\
\midrule
$\text{Re}\,a_{\overline{K} N}^{(1)}$&$0.21$&$0.24^{+0.001}_{-0.001}$&$-0.02$&$0.43^{+0.001}_{-0.001}$&$0.40$&
$0.40^{+0.17}_{-0.19}$&$0.16^{+0.39}_{-0.44}$&\\
$\text{Re}\,a_{\overline{K} N}^{(0)}$&$0.63$&$0.58^{+0.010}_{-0.010}$&$0.46$&$1.67^{+0.010}_{-0.010}$&$1.50$&
$1.52^{+0.22}_{-0.29}$&$1.11^{+0.47}_{-0.59}$&\\
\midrule
$\text{Re}\,a_{\overline{K} \Sigma}^{(3/2)}$&$-0.23$&$0.25^{+0.001}_{-0.001}$&$-0.29$&$-0.26^{+0.001}_{-0.001}$
&$-0.24$&$-0.24^{+0.20}_{-0.20}$&$-0.33^{+0.44}_{-0.47}$&\\
$\text{Re}\,a_{\overline{K} \Sigma}^{(1/2)}$&$0.45$&$0.36^{+0.016}_{-0.016}$&$0.63$&$1.44^{+0.016}_{-0.016}$
&$1.30^{+0.01}_{-0.03}$&$1.28^{+0.27}_{-0.29}$&$0.98^{+0.59}_{-0.59}$&\\
\midrule
$a_{\overline{K} \Xi}^{(1)}$&$-0.47$&$0.45^{+0.006}_{-0.006}$&$-0.31$&$-0.32^{+0.006}_{-0.006}$
&$-0.32$&$-0.33^{+0.11}_{-0.11}$&$-0.26^{+0.34}_{-0.34}$&\\
$a_{\overline{K} \Xi}^{(0)}$&$0.00$&$0.22^{+0.016}_{-0.016}$&$0.43$&$0.64^{+0.016}_{-0.016}$
&$0.57^{+0.02}_{-0.04}$&$0.48^{+0.30}_{-0.29}$&$0.00^{+0.78}_{-0.68}$&\\
\midrule
$a_{\overline{K} \Lambda}$&$0.00$&$0.36^{+0.006}_{-0.006}$&$-0.11$&$0.24^{+0.006}_{-0.006}$
&$0.17^{+0.06}_{-0.06}$&$0.32^{+0.18}_{-0.19}$&$0.04^{+0.55}_{-0.56}$&\\
\midrule
$\text{Re}\,a_{\eta N}$&$0.00$&$-0.17^{+0.001}_{-0.001}$&$0.11$&$-0.06^{+0.001}_{-0.001}$
&$0.18^{+0.07}_{-0.07}$&$0.31^{+0.23}_{-0.25}$&$0.13^{+0.60}_{-0.65}$&\\
\midrule
$\text{Re}\,a_{\eta \Sigma}$&$0.00$&$0.82^{+0.006}_{-0.006}$&$0.09$&$0.91^{+0.006}_{-0.006}$&
$0.42^{+0.04}_{-0.04}$&$0.25^{+0.08}_{-0.08}$&$0.03^{+0.24}_{-0.24}$&\\
\midrule
$\text{Re}\,a_{\eta \Xi}$&$0.00$&$0.71^{+0.014}_{-0.014}$&$0.02$&$0.72^{+0.014}_{-0.014}$&
$0.66^{+0.08}_{-0.08}$&$0.73^{+0.26}_{-0.27}$&$0.25^{+0.74}_{-0.73}$&\\
\midrule
$\text{Re}\,a_{\eta \Lambda}$&$0.00$&$0.16^{+0.006}_{-0.006}$&$0.11$&$0.27^{+0.006}_{-0.006}$
&$0.69^{+0.11}_{-0.11}$&$0.32^{+0.16}_{-0.19}$&$0.15^{+0.51}_{-0.55}$&\\
\bottomrule
\midrule
\end{tabular}
\begin{tablenotes}
\item[1]The scattering lengths are calculated by threshold T-matrices in the framework of HB$\chi$PT from Ref.~\cite{liu20071}.
\item[2]The scattering lengths are calculated in the framework of HB$\chi$PT from Ref.~\cite{mai2009}.
\item[3]The scattering lengths are calculated in the framework of infrared regularization (IR) from Ref.~\cite{mai2009}.
\end{tablenotes}
\end{threeparttable}
\end{table*}

\appendix\markboth{Appendix}{Appendix}
\renewcommand{\thesection}{\Alph{section}}
\numberwithin{equation}{section}
\section{Loop functions}
\label{One-loop amplitudes}
In this Appendix we present the basic loop-functions:
\begin{align}
\label{A1}
J_0(w,m)=&\frac{1}{i}\int\frac{d^{D}l}{(2\pi)^D}\frac{1}{(v\cdot l-w)(m^2-l^2)}
=\frac{w}{8\pi^2}\Big(1-2\text{ln}\frac{m}{\lambda}\Big)\nonumber\\
&+\begin{cases}
\dfrac{1}{4\pi^2}\sqrt{w^2-m^2}\text{ln}\dfrac{-w+\sqrt{w^2-m^2}}{m}& (w<-m),\\
-\dfrac{1}{4\pi^2}\sqrt{m^2-w^2}\text{arccos}\dfrac{-w}{m}& (-m<w<m),\\
\dfrac{1}{4\pi^2}\sqrt{w^2-m^2}\Bigg(i\pi-\text{ln}\dfrac{w+\sqrt{w^2-m^2}}{m}\Bigg)&(w>m),
\end{cases}
\end{align}
\begin{align}
\label{A2}
\frac{1}{i}\int\frac{d^{D}l}{(2\pi)^{D}}\frac{\{1,l^\mu,l^\mu l^\nu\}}{(m^2-l^2)[m^2-(l-k)^2]}=\{I_0(t,m),\frac{k^\mu}{2}I_0(t,m),g^{\mu\nu}I_2(t,m)+k^\mu k^\nu I_3(t,m)\},
\end{align}
\begin{align}
\label{A3}
I_0(t,m)=\frac{1}{8\pi^2}\Bigg\{\frac{1}{2}-\text{ln}\frac{m}{\lambda}-\sqrt{1-\frac{4m^2}{t}}\text{ln}\frac{\sqrt{4m^2-t}+\sqrt{-t}}{2m}\Bigg\},
\end{align}
\begin{align}
\label{A4}
I_2(t,m)=\frac{1}{48\pi^2}\Bigg\{2m^2-\frac{5t}{12}+\Big(\frac{t}{2}-3m^2\Big)\text{ln}\frac{m}{\lambda}-\frac{(4m^2-t)^{3/2}}{2\sqrt{-t}}\text{ln}\frac{\sqrt{4m^2-t}+\sqrt{-t}}{2m}\Bigg\},
\end{align}
\begin{align}
\label{A5}
I_3(t,m)=\frac{1}{24\pi^2}\Bigg\{\frac{7}{12}-\frac{m^2}{t}-\text{ln}\frac{m}{\lambda}-\Big(1-\frac{m^2}{t}\Big)\sqrt{1-\frac{4m^2}{t}}\text{ln}\frac{\sqrt{4m^2-t}+\sqrt{-t}}{2m}\Bigg\},
\end{align}
\begin{align}
\label{A6}
H_0(t,m_1,m_2)=&\frac{1}{i}\int\frac{d^{D}l}{(2\pi)^D}\frac{1}{v\cdot l(m_1^2-l^2)[m_2^2-(l-k)^2]}\nonumber\\
=&\frac{1}{8\pi\sqrt{-t}}\text{arctan}\frac{\sqrt{-t}}{m_1+m_2},
\end{align}
where $t=k\cdot k<0$. Note that terms proportional to the divergent constant $\lambda^{
D-4}[\frac{1}{D-4}+{\frac{1}{2}}(\gamma_E-1-\ln 4\pi)]$ have been dropped. 

The explicit analytical expression for the loop contributions of order $\mathcal{O}(q^3)$ to the meson-baryon scattering amplitudes can be obtained from the first author upon request. 

\bibliographystyle{apsrev4-1}
\bibliography{latextemplate}

\end{document}